\documentclass[a4paper,fleqn,usenatbib]{mnras}
\usepackage{newtxtext,newtxmath}
\usepackage[T1]{fontenc}
\usepackage{ae,aecompl}
\pdfoutput=1
\usepackage{graphicx}	
\usepackage{amsmath}	
\usepackage{amssymb}	
\usepackage{multicol}  
\usepackage{threeparttable}

\title[A distant QSO--starburst system caught by ALMA]{SMM J04135+10277: A distant QSO--starburst system caught by ALMA}
\author[J. Fogasy et al.]{
Judit Fogasy,$^{1}$\thanks{E-mail: judit.fogasy@chalmers.se}
K. K. Knudsen,$^{1}$
G. Drouart,$^{2}$
C. D. P. Lagos,$^{3,4,5}$
and L. Fan$^{6,7}$
\\
$^{1}$Chalmers University of Technology, Department of Space, Earth and Environment, Onsala Space Observatory, 439 92 Onsala, Sweden\\
$^{2}$Curtin University, Department of Physics and Astronomy, Kent Street, Bentley, 6102, Perth, WA, Australia\\
$^{3}$International Centre for Radio Astronomy Research (ICRAR), M468, University of Western Australia, 35 Stirling Hwy, Crawley, WA 6009, Australia\\
$^{4}$ARC Centre of Excellence for All Sky Astrophysics in 3 Dimensions (ASTRO 3D)\\
$^{5}$Cosmic Dawn Center (DAWN), Copenhagen, Denmark\\
$^{6}$CAS Key Laboratory for Research in Galaxies and Cosmology, Department of Astronomy, University of Science and Technology of China, Hefei 230026, China\\
$^{7}$School of Astronomy and Space Sciences, University of Science and Technology of China, Hefei, Anhui 230026, People's Republic of China
}

\date{Accepted XXX. Received YYY; in original form ZZZ}

\pubyear{2019}

\begin{document}
\label{firstpage}
\pagerange{\pageref{firstpage}--\pageref{lastpage}}
\maketitle

\begin{abstract}
The gas content of galaxies is a key factor for their growth, starting from star formation and black hole accretion to galaxy mergers. Thus, characterising its properties through observations of tracers like the CO emission line is of big importance in order to understand the bigger picture of galaxy evolution.
We present Atacama Large Millimeter/submillimeter Array (ALMA) observations of dust continuum, CO(5-4) and CO(8-7) line emission in the quasar--star-forming companion system SMM J04135+10277 ($z=2.84$). Earlier low-$J$ CO studies of this system found a huge molecular gas reservoir associated to the companion galaxy, while the quasar appeared gas-poor. Our CO observations revealed that the host galaxy of the quasar is also gas-rich, with an estimated molecular gas mass of $\sim(0.7-2.3)\times10^{10}\,\rm M_{\sun}$. 
The CO line profiles of the companion galaxy are very broad ($\sim1000\,\rm km\,s^{-1}$), and show signs of rotation of a compact, massive system. 
In contrast to previous far-infrared observations, we resolve the continuum emission and detect both sources, with the companion galaxy dominating the dust continuum and the quasar having a $\sim25\%$ contribution to the total dust emission. 
By fitting the infrared spectral energy distribution of the sources with \textsc{MR-MOOSE} and empirical templates, the infrared luminosities of the quasar and the companion are in the range of $L_{\rm IR, QSO}\sim(2.1-9.6)\times10^{12}\,\rm L_{\sun}$ and $L_{\rm IR, Comp.}\sim(2.4-24)\times10^{12}\,\rm L_{\sun}$, while the estimated star formation rates are $\sim210-960\,\rm M_{\sun}\,yr^{-1}$ and $\sim240-2400\,\rm M_{\sun}\,yr^{-1}$, respectively.
 Our results demonstrate that non-detection of low-$J$ CO transition lines in similar sources does not necessarily imply the absence of massive molecular gas reservoir but that the excitation conditions favour the excitation of high-$J$ transitions.
\end{abstract}




\section{Introduction}
%
An intense phase of star formation and supermassive black hole (SMBH) growth at high-redshift is necessary to explain the observed correlation found in local elliptical galaxies between the mass of the SMBH and some properties of the host galaxy, such as the velocity dispersion or bulge mass \citep[e.g.][]{1998AJ....115.2285M, 2000ApJ...539L..13G, 2001ApJ...547..140M, 2002ApJ...574..740T, 2003ApJ...589L..21M, 2004ApJ...604L..89H, 2013ApJ...764..184M}. The found correlations suggest that the evolution of the SMBHs and their host galaxies is tied together. Theoretical studies of galaxy evolution provide a possible framework for this co-evolution in form of interactions and major merger events between massive and gas-rich galaxies \citep[e.g.][]{1988ApJ...325...74S, 2005Natur.433..604D, 2008A&A...492...31D, 2008ApJS..175..356H, 2006ApJS..163....1H, 2010MNRAS.407.1701N}.
As the star formation and black hole activity peaks at around $z\sim2-3$ \citep[e.g.][]{1998ApJ...498..106M, 2004ApJ...615..209H, 2006AJ....131.2766R, 2009ApJ...707.1566Z, 2010MNRAS.401.2531A, 2015MNRAS.451.1892A}, by probing the far-infrared (FIR) properties and molecular gas content of high-$z$ active galactic nuclei (AGNs) and their close environment, we can probe this scenario. 
Moreover, there is an extensive literature with ambigous conclusions on whether there is a strong correlation between the star formation rate (SFR) and AGN luminosity of high-redshift AGNs 
\citep[e.g.][]{2006ApJ...649...79S, 2010ApJ...712.1287L, 2011MNRAS.416...13B, 2012MNRAS.419...95M, 2012A&A...540A.109S, 2015MNRAS.453..591S, 2016MNRAS.460..902B, 2016ApJ...824...70D, 2016ApJ...819..123N, 2016MNRAS.462.4067P, 2017A&A...604A..67D, 2019MNRAS.486.4320R, 2019MNRAS.488.1180S}. 
Based on these studies the growth of AGNs can follow two different paths: low or moderate luminosity AGNs can evolve through secular processes, thus the SMBH growth is not directly linked to the star formation of the host galaxy; while the most luminous AGNs grow through major mergers which enhance the star formation of the hosts.

Indeed, in the last decade several quasar studies at submm wavelength found signs of intense star formation in their hosts by tracing dust continnum emission and \ion{[C}{ii]} cooling line emission \citep[e.g.][]{2001ApJ...555..625C, 2001A&A...374..371O, 2005A&A...440L..51M, 2013ApJ...773...44W, 2016ApJ...816...37V}. However, most of these studies are limited in resolution and/or sensitivity, thus they could miss the detection of close companion galaxies, which also have a contribution to the submm emission.\\
\indent
Thanks to the high angular resolution and sensitivity of the Atacama Large Millimeter/submillimeter Array (ALMA), many recent studies of high-$z$ quasars found companion galaxies in their vicinity, in some cases more than one \citep[e.g.][]{2012ApJ...752L..30W, 2013ApJ...763..120C, 2015ApJ...806L..25S, 2017ApJ...836....8T, 2017MNRAS.465.4390B, 2018MNRAS.479.1154B, 2017A&A...605A.105C, 2017Natur.545..457D, 2018ApJ...856L...5F}. These results are in agreement with our earlier study, where we used \textsc{galform} \citep{2014MNRAS.439..264G, 2014MNRAS.440..920L}, a galaxy formation and evolution model with the aim of investigating the expected frequency of finding quasar--companion galaxy systems \citep{2017A&A...597A.123F}.\\
\indent
In this paper we focus on the quasar--star-forming companion galaxy system SMM J04135+10277 ($z=2.84$). The AGN component of the system is a type-1 quasar, originaly discovered at submm wavelength \citep{2003A&A...411..343K} using the Submillimetre Common-User Bolometer Array (SCUBA).  Subsequent single-dish observations of the quasar tracing low-$J$ CO transitions revealed a massive molecular gas reservoir associated with the AGN ($M_{\rm H_{2}}\sim 10^{11} \rm M_{\sun}$; \citealt{2004ApJ...609...61H, 2011ApJ...739L..32R}). However, follow-up interferometric observations showed that the gas reservoir was offset from the position of the quasar  by $\sim 5 \arcsec$ (41.5 kpc) and associated with a close companion galaxy \citep{2013ApJ...765L..31R}. Both the quasar and its companion are slightly gravitationally lensed by the foreground galaxy cluster Abell 478, by a gravitational magnification factor of $\mu_{\rm QSO}=1.3$ and $\mu_{\rm Comp.}=1.6\pm0.5$, respectively \citep{2003A&A...411..343K, 2013ApJ...765L..31R}. Given the molecular gas distribution and the separation between the sources, the system has been proposed to be in an early stage of a wet-dry merger event, the companion being gas-rich and the quasar being gas-poor, making this case unique \citep{2013ApJ...765L..31R}.
Based on the spectral energy distribution (SED) of the companion galaxy we can infer that it is a heavily dust-obscured star-forming galaxy ($A_{\rm V}\sim2.8\,\rm mag$), with a star formation rate of $\sim700\,\rm M_{\sun}\,yr^{-1}$ and dust mass of $\sim5\times 10^{9}\,\rm M_{\sun}$ \citep{2017A&A...597A.123F}. Considering the high SFR, molecular gas and dust mass of the companion galaxy, it is possible that the companion has a significant contribution to the SCUBA detected submm emission and the majority of star formation happens in the companion rather than in the host galaxy of the quasar.

In order to further investigate this system we obtained dust continuum, CO(5-4) and CO(8-7) observations with ALMA. Our goal is to constrain the excitation properties of SMM J04135+10277 through modelling of the spectral line energy distribution (SLED) of both sources and to deepen our understanding of the processes shaping the evolution of this system. This is important because prior to ALMA, the majority of studies finding AGN--companion systems were focusing on the large scale environment of AGNs and did not study the systems in details. 

In section \ref{sec:observations} we present the ALMA observations, in section \ref{sec:analysis} we describe the results of the observations and the data analysis, including the SLED modelling of the sources. In section \ref{sec:discussion} we discuss our findings and compare them with other studies found in the literature. Finally, we summarise our results in section \ref{sec:conclusion}.
Throughout this paper we adopt WMAP7 cosmology with $H_{0}=70.4\ \rm{km \ s^{-1} Mpc^{-1}}$, $\rm \Omega_{m}=0.272$ and $\rm \Omega_{\Lambda}=0.728$ \citep{2011ApJS..192...18K}.
%
\section{Observations}
\label{sec:observations}

\begin{table}
	\centering
	\caption{Details of the ALMA observations of SMM J04135+10277.}
	\label{tab:table_obs}
	\begin{tabular}{lcc}
		\hline
		\hline
		 & Band 4& Band 6\\
		\hline
		Date & 2016-06-02& 2016-03-31\\
		Configuration & C40-4 & C36-2/3\\
		$\rm N_{antennas}$ & 37 & 44\\
		$\rm T_{\rm exp}$ (min) & 10.6& 10.1\\
		Beam size ($\arcsec$)& $1.01\times0.92$& $1.0\times0.78$\\
		$\rm RMS_{cont.}\,\rm (\mu Jy\,beam^{-1})$ & 27.8 & 46.7\\
		$\rm RMS_{line}\,\rm ( mJy\,beam^{-1})$\textsuperscript{a}& $0.32$ & $0.38$\\
		\hline
		 \multicolumn{3}{l}{Notes}\\
		 \multicolumn{3}{l}{\footnotesize{$^a$ The spectra are binned to $77\,\rm km\,s^{-1}$ in each band.}}\\
	\end{tabular}
\end{table}

SMM J04135+10277 was observed with the 12-m ALMA array during cycle 3 (project code 2015.1.00661.S; P.I.: Fogasy).
One of the spectral windows with a bandwidth of 1.875 GHz was tuned to the central frequency of 149.843 GHz to cover the redshifted CO(5-4) line in band 4, and to 239.648 GHz to cover the redshifted CO(8-7) line in band 6.
The other three spectral windows of each band, with a bandwidth of 2 GHz each, were used to observe the continuum. Other details of the observations are given in Table \ref{tab:table_obs}. 

The calibration of the data was done using the ALMA Science Pipeline. The calibration process includes standard calibration and reduction steps, such as flagging, bandpass calibration, flux and gain calibration. The quasar J0423-0120 was used as a bandpass calibrator source. Due to the strong variability of the flux calibrator source at the time of the observations ($\sim10\%$ in both bands), the flux calibration was redone with updated flux values. 
After this correction we assume a conservative 10\%  absolute flux calibration error, which is not included in the flux densities reported in this paper.

To image the data we used the Common Astronomy Software Applications (CASA; \citealt{2007ASPC..376..127M}) with the ARC Node provided scripts.
We applied the \textsc{CLEAN} algorithm to create the continuum images and line data cubes using natural weighting.
In case of the spectral line data the continuum was fitted using the line free channels of every spectral window and was subtracted in the $uv$-plane using the \textsc{uvcontsub} task in CASA. For each band the continuum subtracted line data was imaged with $77\,\rm km\,s^{-1}$ channel width.

\section{Results and analysis}
\label{sec:analysis}
\subsection{Dust continuum emission, FIR luminosities and star formation rates}
\begin{figure}
	\centering
	\includegraphics[width=0.9\columnwidth]{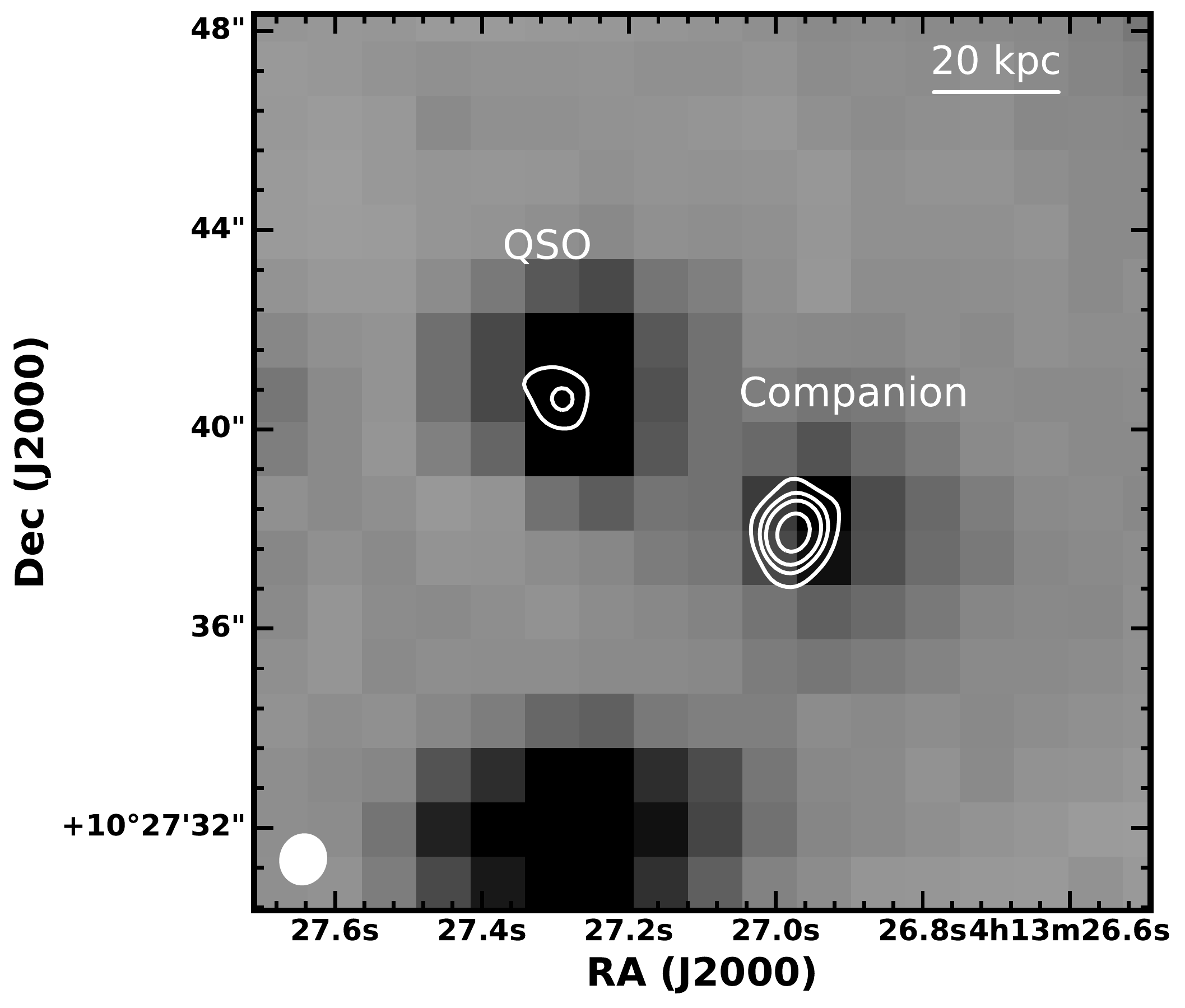}
	\includegraphics[width=0.9\columnwidth]{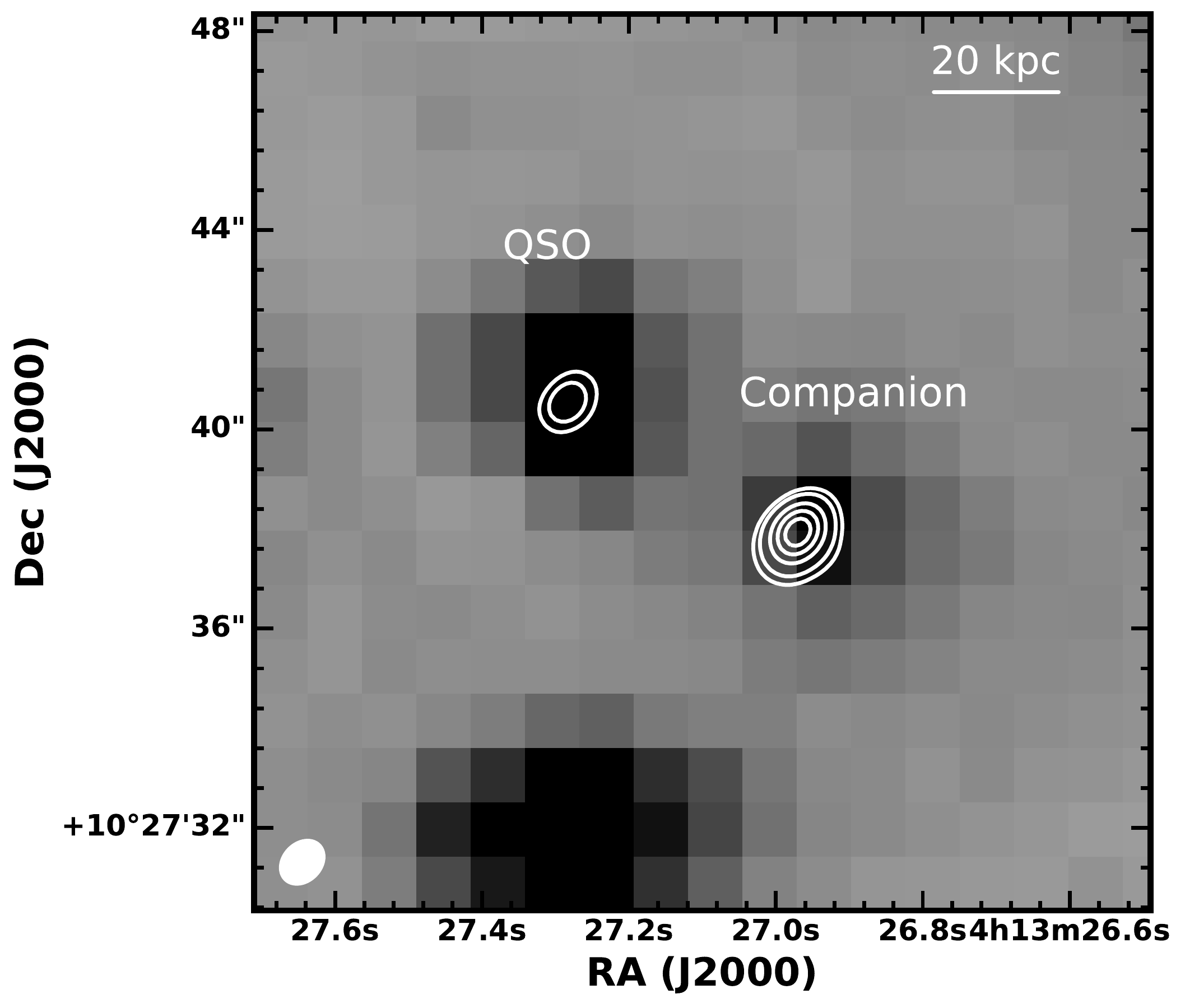}
	\caption{ALMA 150 GHz (\textit{upper panel}) and 240 GHz (\textit{lower panel}) dust continuum contours overlayed on the \textit{Spitzer} IRAC $4.5\,\mu\rm m$ image of SMM J04135+10277. The beam size is indicated in the bottom left corner. The contour levels are $5\sigma$, $10\sigma$, $15\sigma$, $25\sigma$ for the 150 GHz continuum and $25\sigma$, $50\sigma$, $75\sigma$, $100\sigma$ for the 240 GHz continuum.}
	\label{fig:continuum_Irac}
\end{figure}

\begin{table}
	\centering
	\caption{Positions, continuum fluxes and physical source sizes of the system of SMM J04135+10277.}
	\label{tab:table_cont}
	\begin{tabular}{lcc}
		\hline
		\hline
		& QSO & Companion\\
		\hline
		R.A. &04:13:27.28 & 04:13:26.98 \\
		Dec. & +10:27:40.41& +10:27:37.89\\
		$S_{150\,\rm GHz}$ (mJy) &$0.399\pm0.015$ & $1.233\pm0.037$ \\
		$S_{240\,\rm GHz}$ (mJy) & $2.174\pm0.023$& $7.64\pm0.30$\\
		Size at 150 GHz ($\arcsec$) & $0.67\times 0.32$& $0.71\times 0.44$\\
		Size at 240 GHz ($\arcsec$)&$0.53\times 0.37$ & $ 0.65\times 0.45$\\
		\hline
	\end{tabular}
\end{table}
The ALMA observations enabled us to resolve the far-IR emission of SMM J04135+10277 and detect both sources for the first time (Fig.\ref{fig:continuum_Irac}), in contrast to previous single-dish submm observations, where the quasar and its companion remained unresolved within a single beam (e.g. SCUBA beam is $15\arcsec$ at $850\,\mu\rm m$, the separation between the quasar and the companion is $\sim5\arcsec$). Thanks to the high-sensitivity of ALMA we achieved a robust detection, yielding a high signal-to-noise ratio of 14 and 46 in the case of the quasar and 44 and 163 in the case of the companion galaxy, in band 4 and band 6, respectively.
The dust emission of the quasar is much weaker compared to the companion galaxy, only about $22-25\%$ of the total emission in both bands (Table \ref{tab:table_cont}). 
In band 4 the cD galaxy of A478 is detected north-west from the phase centre $(\rm R.A.=04^{h}13^{m}25^{s}.27,\,Dec.=+10^{\degr}27^{'}54^{''}.69$), at $\sim33\arcsec$ ($\sim 264\,\rm kpc$) distance from the quasar but it is not detected in the band 6 data.

\begin{figure*}
	\centering
	\includegraphics[width=0.98\textwidth]{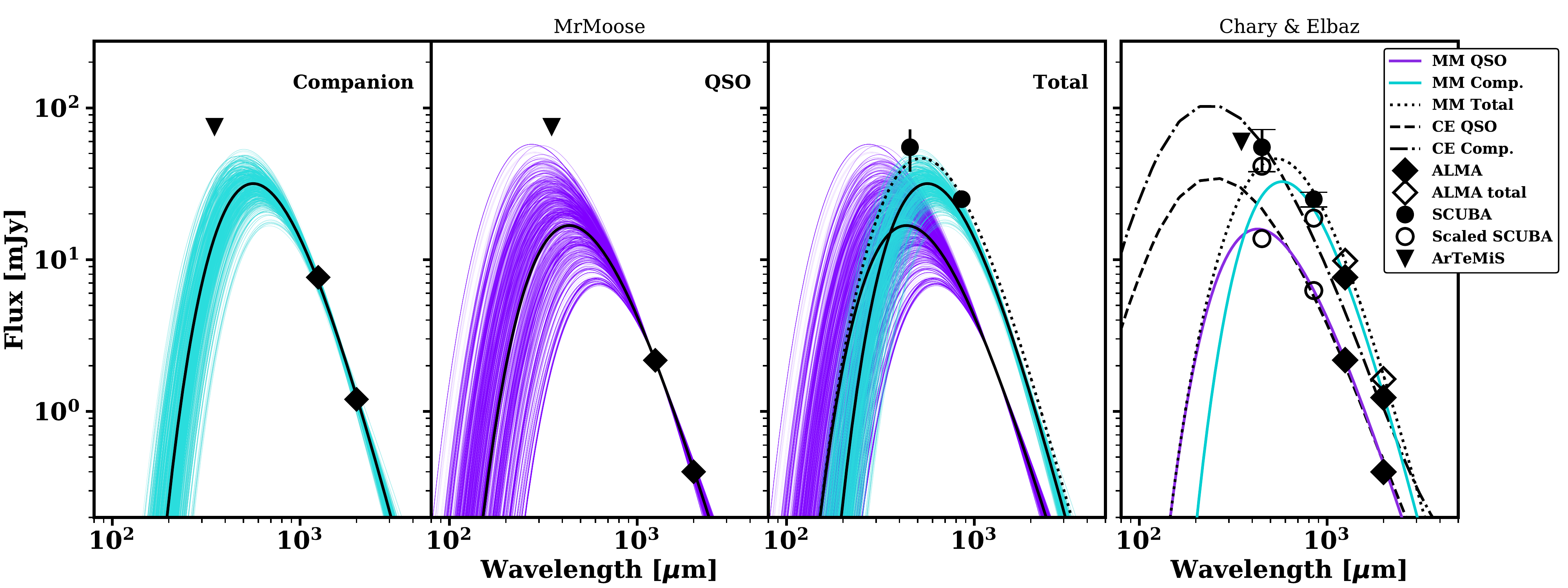}
	\caption{The FIR SED of the quasar and its companion. The first three panels show the \textsc{MR-MOOSE} modified blackbody SED of the companion, the quasar and the total emission, respectively. The purple and turquoise lines show the possible range of solutions. The third panel shows the \textsc{MR-MOOSE} best fit models of the quasar and the companion (solid lines), and the total emission (dotted line). The fourth panel shows the best-fit \textsc{MR-MOOSE} SEDs as turquoise, purple and dotted lines, and the best-fit \citet{2001ApJ...556..562C} SED of the quasar and the companion (dashed and dash-dotted lines).  On all panels the symbols represent the same observations. The black diamonds represent the ALMA observations; the black unfilled diamonds show the total ALMA emission of the quasar and the companion. The black circles show the total SCUBA $450\,\mu\rm m$ and $850\,\mu\rm m$ emission; the black unfilled circles represent the SCUBA observations, scaled by the assumed relative contribution of the quasar (25\% ) and companion (75\% ) to the total emission. The upper limit shows the $3\sigma$ ArTeMiS measurement of the system.}
	\label{fig:mrmooseSED}
\end{figure*}

The UV-to-IR SED of the companion galaxy was constructed in \citet{2017A&A...597A.123F}, here we focus on the FIR SED of the quasar and the companion. 
To complement our ALMA observations we used previous SCUBA and ArTeMiS data to obtain the SED \citep{2003A&A...411..343K, 2017A&A...597A.123F}. As the resolution of the single-dish data is not sufficient to separate emission associated to the quasar and the companion
galaxy, we use the multi-wavelength SED fitting code \textsc{MR-MOOSE} \citep{2018MNRAS.477.4981D}. The code is designed to treat upper limits consistently as well as large variations of resolution and to fit blended sources
using a Bayesian framework, by allowing the user to use the spatial information. \textsc{MR-MOOSE} simultaneously fits the SEDs of the quasar and the companion, and the total emission when it is impossible
to resolve sources individually, e.g. in the SCUBA bands. The best fit of models and the possible range of solutions are shown on Figure \ref{fig:mrmooseSED}. In the case of the quasar the dust emissivity and dust temperature of the best-fit models are $\beta_{\rm dust, QSO}=2.5\pm0.3$ and $T_{\rm dust, QSO}=24\pm6\,\rm K$. In the case of the companion the best-fit models predicts $\beta_{\rm dust, Comp.}=3.3\pm0.4$ and $T_{\rm dust, Comp.}=16\pm3\,\rm K$. 
Considering the small number of available data points, it is difficult to properly constrain the absolute values of $\beta$ and $T_{\rm dust}$, but the \textsc{MR-MOOSE} result still highlights that the quasar has a consistently higher temperature compared to the companion.

The FIR luminosities of the quasar and its companion based on the best-fit \textsc{MR-MOOSE} SED models are $L_{\rm FIR, QSO}=(2.1^{+1.7}_{-1.2})\times10^{12}\,\rm L_{\sun}$ and  $L_{\rm FIR, Comp.}=(2.4\pm1)\times10^{12}\,\rm L_{\sun}$, corrected for gravitational lensing magnification. From the obtained infrared luminosities, the SFR of the sources can be estimated according to the relation $\rm SFR\sim\delta_{MF}\times10^{-10}\,\it L_{\rm IR}$ \citep{2013ARA&A..51..105C}, assuming that the dust is heated by star formation. As  \textsc{MR-MOOSE} only fits the FIR part of the SED, the total IR luminosity of the sources could be higher. Thus we treat the SFR calculated from the $L_{\rm FIR}$ as a lower limit. By adopting a Chabrier initial mass function ($\delta_{MF}\sim1$), we get a SFR of $\sim210\,\rm M_{\sun}\,yr^{-1}$ and $\sim240\,\rm M_{\sun}\,yr^{-1}$ for the quasar and the companion, respectively.

In addition, we used the empirical SED model grid of \citet{2001ApJ...556..562C} as an alternative approach to fit the IR SED of quasar and the companion. The grid includes 105 templates with a wide range of infrared luminosities. As the SED templates have no free parameters, we chose the templates that fit our ALMA detections the best. For the SCUBA bands we assumed a $25\%$ contribution from the quasar, based on the ALMA detections. The best-fit templates are shown on Figure \ref{fig:mrmooseSED} together with the \textsc{MR-MOOSE} models. Based on the best-fit \citet{2001ApJ...556..562C} templates the IR luminosities of the quasar and the companion are $L_{\rm IR, QSO}\sim9.6\times10^{12}\,\rm L_{\sun}$ and  $L_{\rm IR, Comp.}\sim2.4\times10^{13}\,\rm L_{\sun}$ (the luminosities are corrected for the gravitational lensing magnification). The SFR of the quasar and its companion according the same relation as above is $\sim960\,\rm M_{\sun}\,yr^{-1}$ and $\sim2400\,\rm M_{\sun}\,yr^{-1}$, respectively.

By comparing the results of the two approaches, the \citet{2001ApJ...556..562C} templates cannot fit well the data unless the quasar's contribution at the $450\,\micron$ SCUBA wavelength is higher ( $\sim35\%$). Furthemore, the \citet{2001ApJ...556..562C} templates are constructed for low-redshift star-forming galaxies and might not be suitable for high-$z$ AGNs. As \textsc{MR-MOOSE} only fits a modified-blackbody SED to the FIR part of the SED and the \citet{2001ApJ...556..562C} templates cover the whole luminosity, the results of these two approaches can be treated as lower and upper limits for the total IR luminosity and the SFR of the sources.

%
\subsection{CO(5-4) and CO(8-7) line emission}

\begin{figure*}
	\centering
	\includegraphics[width=1.0\columnwidth]{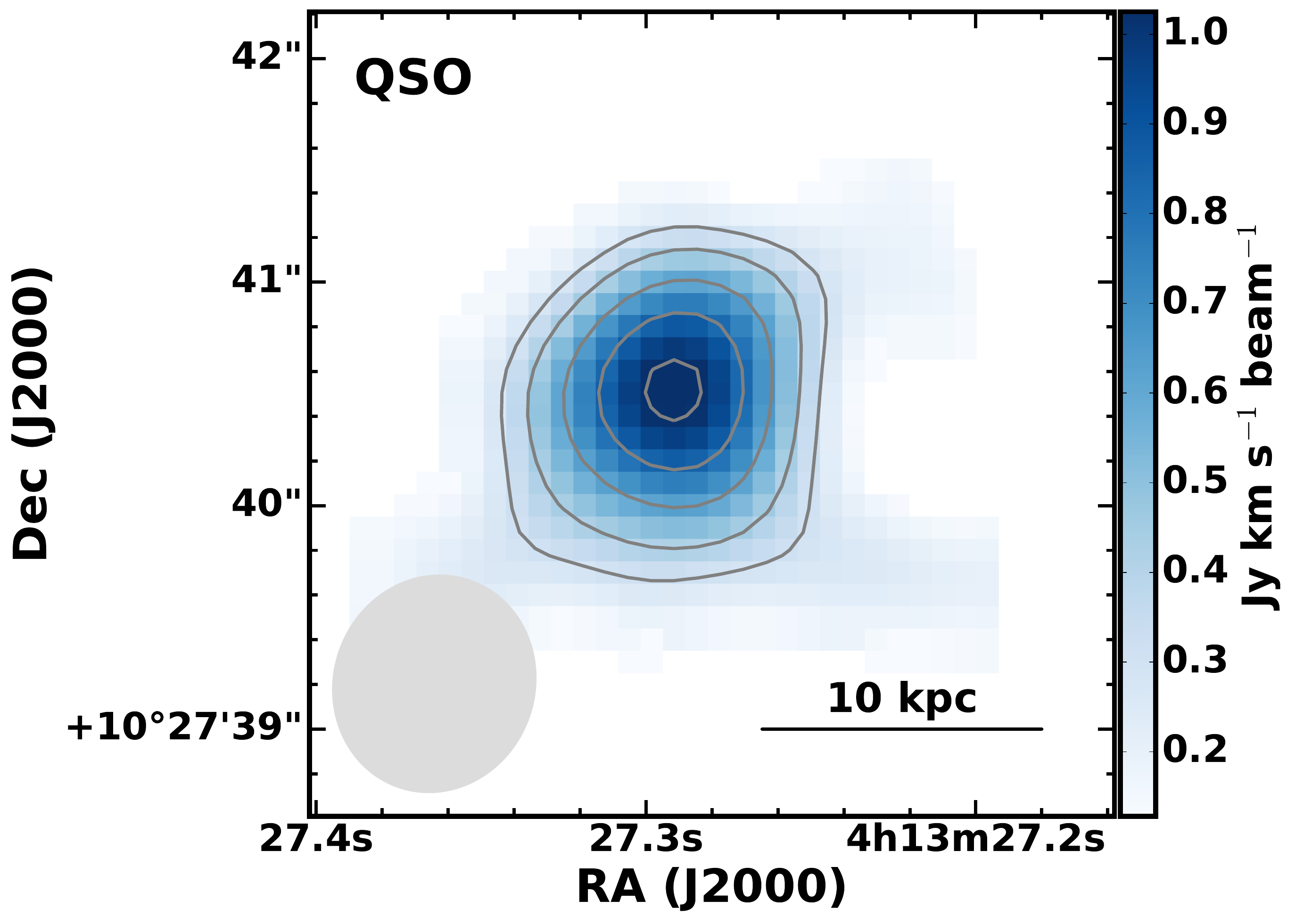}
	\includegraphics[width=1.0\columnwidth]{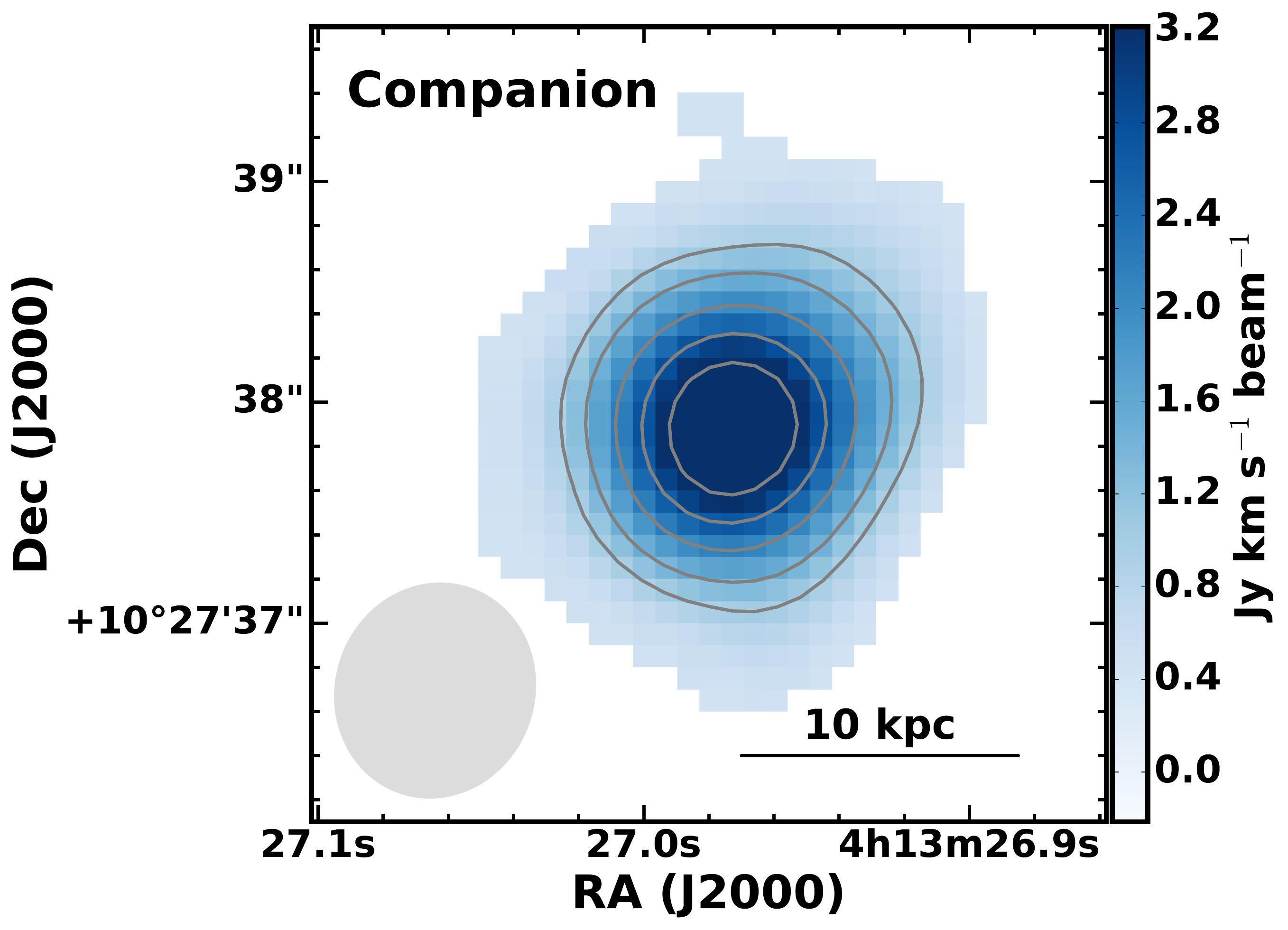}
	\includegraphics[width=1.0\columnwidth]{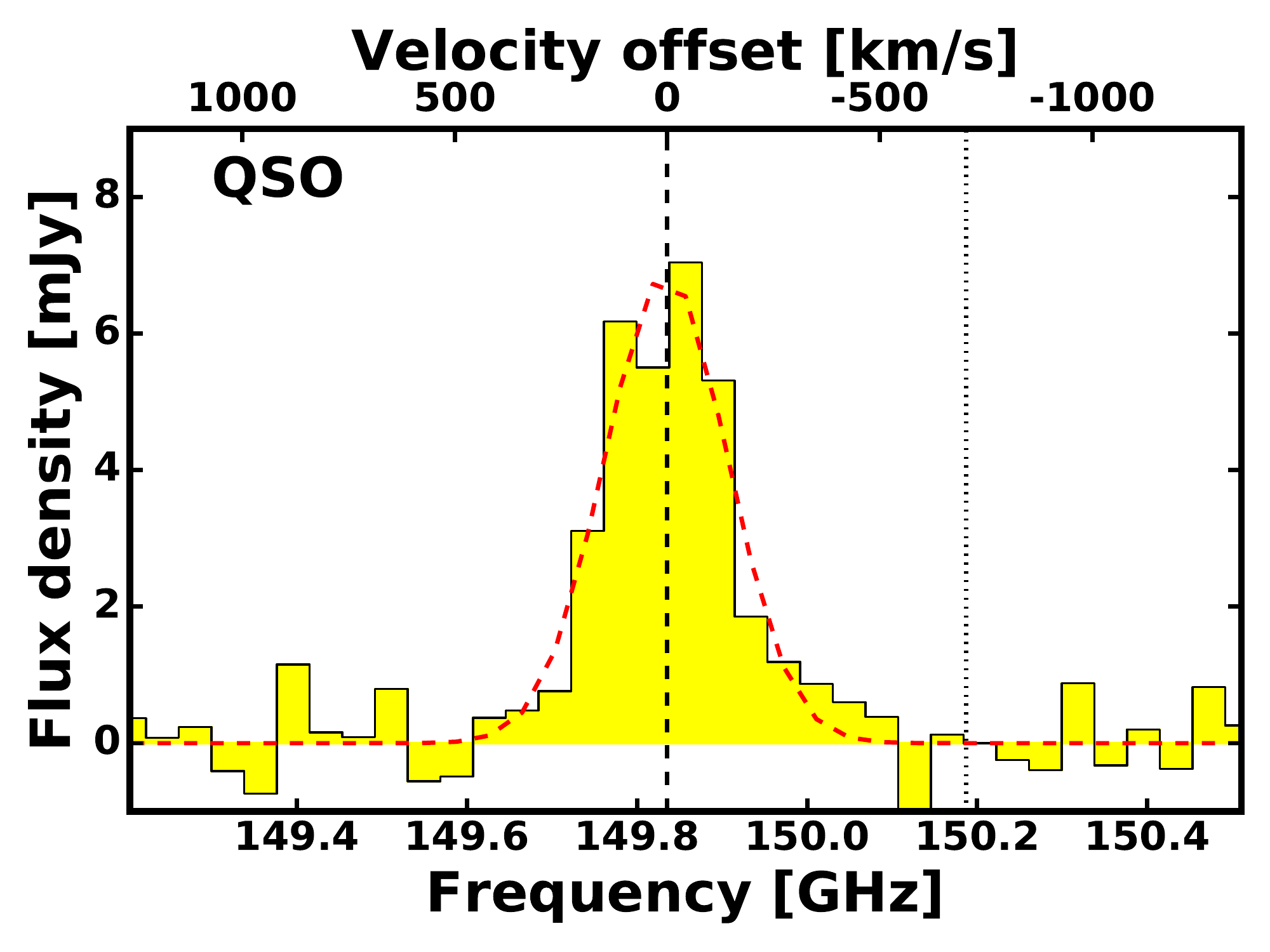}
	\includegraphics[width=1.0\columnwidth]{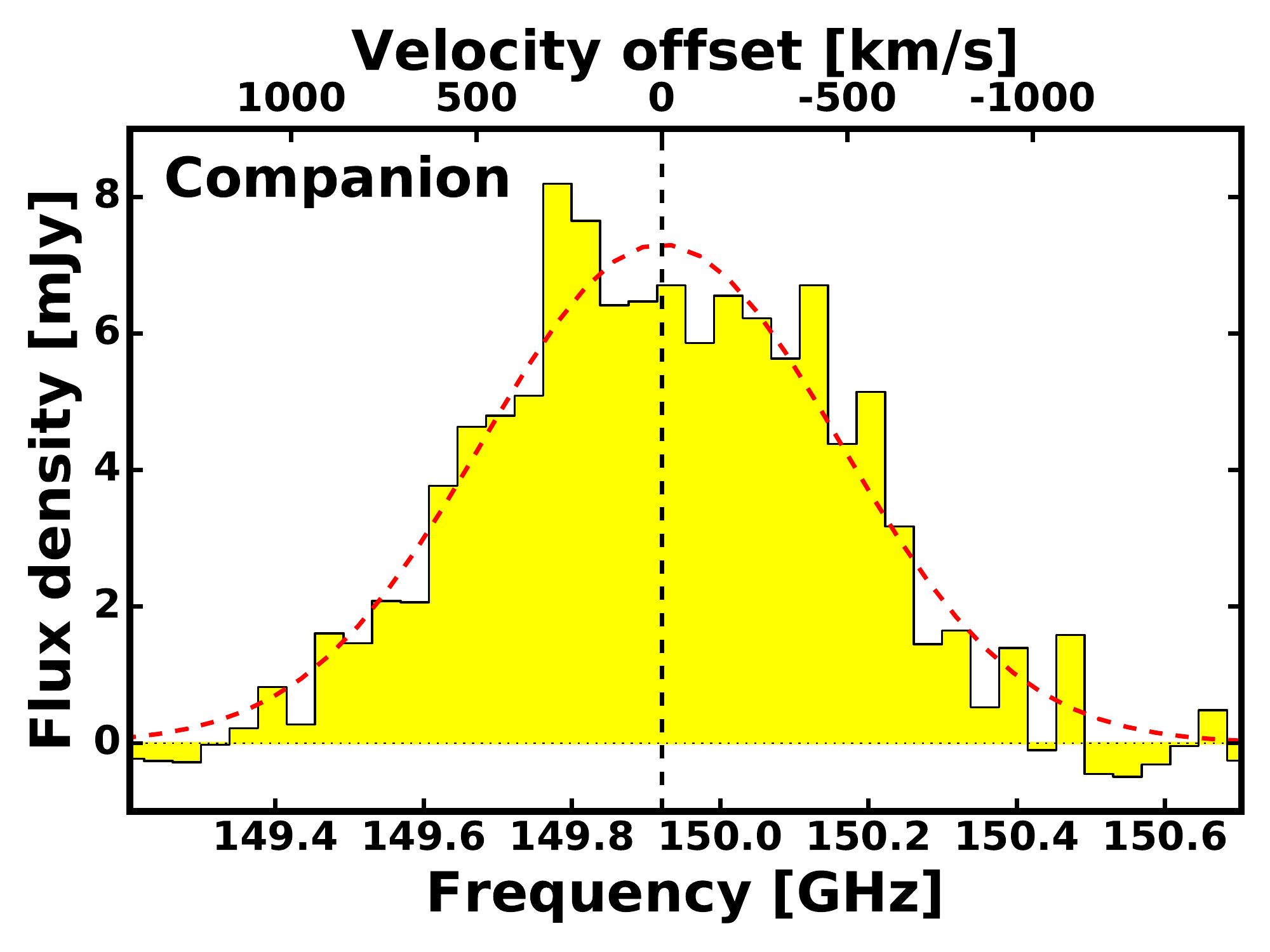}
	\caption{CO(5-4) emission of the system of SMM J04135+10277. \textit{Top:} Integrated intesity contour maps of the quasar and its companion. The contour levels are at $[5,7,10,13,16]\times\sigma$, where $\sigma=0.064\rm\,Jy\,beam^{-1}\,km\,s^{-1}$ for the QSO and $\sigma=0.214\rm\,Jy\,beam^{-1}\,km\,s^{-1}$ for the companion. ALMA beams are shown as grey ellipses at the bottom left corner.
\textit{Bottom:} The CO(5-4) spectra of SMM J04135+10277 extracted at the position of the quasar and the companion galaxy, respectively. The spectra are binned to $77\,\rm km\,s^{-1}$ per channel and the continuum is subtracted. The red curves show the single component Gaussian fits to the line profiles. The vertical dashed lines indicate the redshifts of the quasar and the companion obtained from the line fitting of the CO(5-4) emission lines. The vertical dotted line indicate the optical redshifts of the quasar ($z=2.837$; \citealt{2003A&A...411..343K}). The top-axis shows the relative velocity offset with respect to the fitted redshifts.}
	\label{fig:band4_line}
\end{figure*}

\begin{figure*}
	\centering
	\includegraphics[width=1.0\columnwidth]{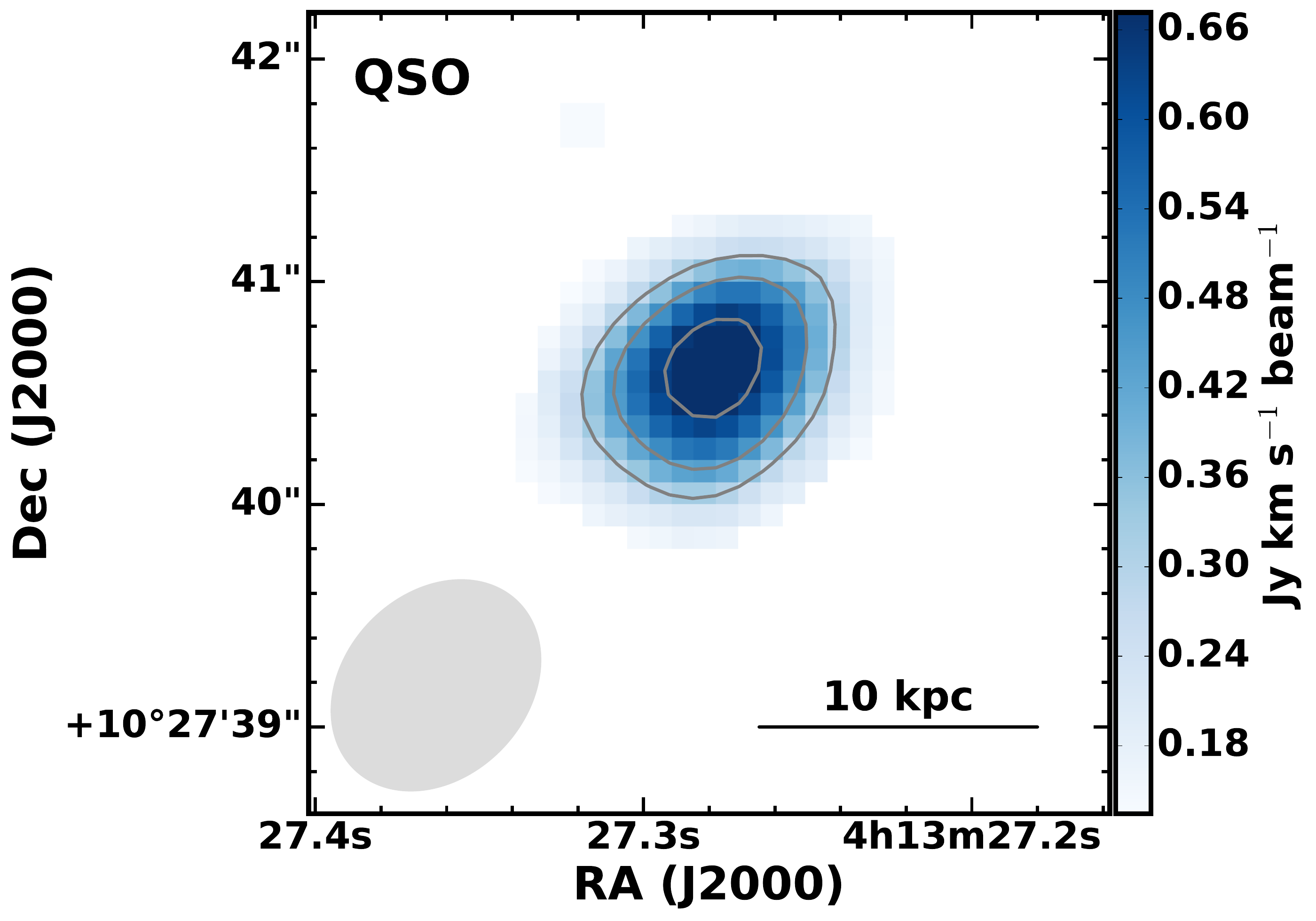}
	\includegraphics[width=1.0\columnwidth]{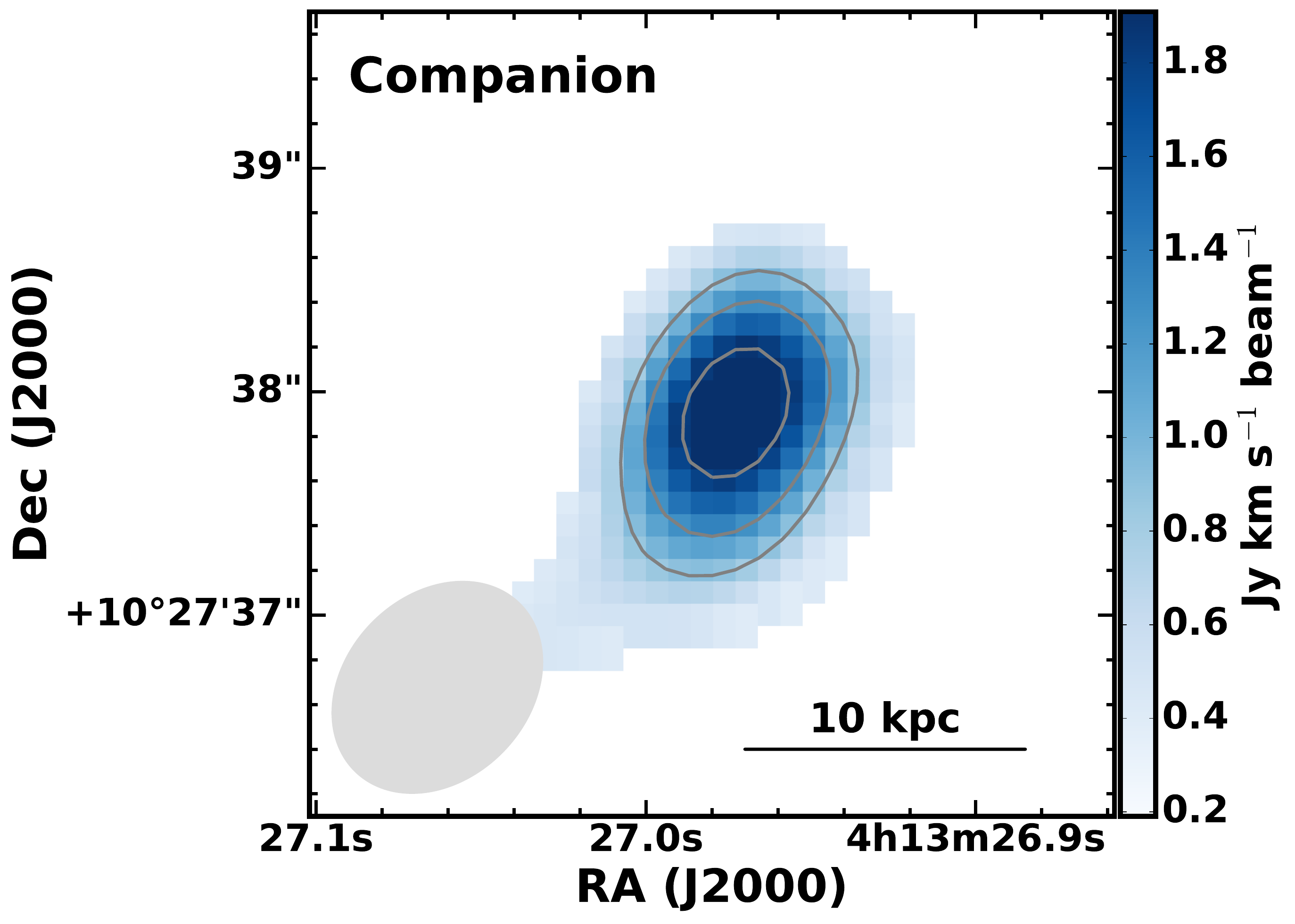}
	\includegraphics[width=1.0\columnwidth]{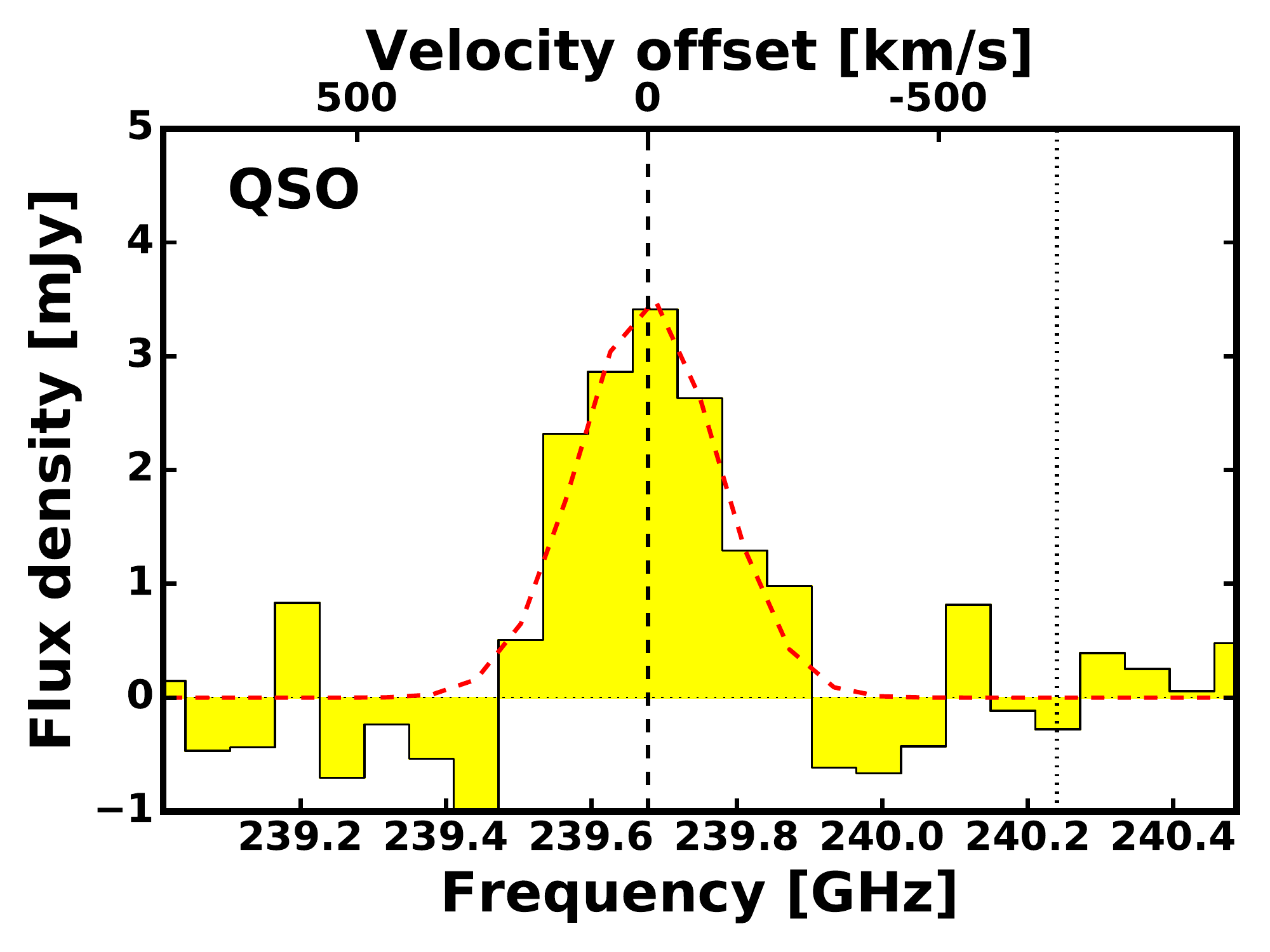}
	\includegraphics[width=1.0\columnwidth]{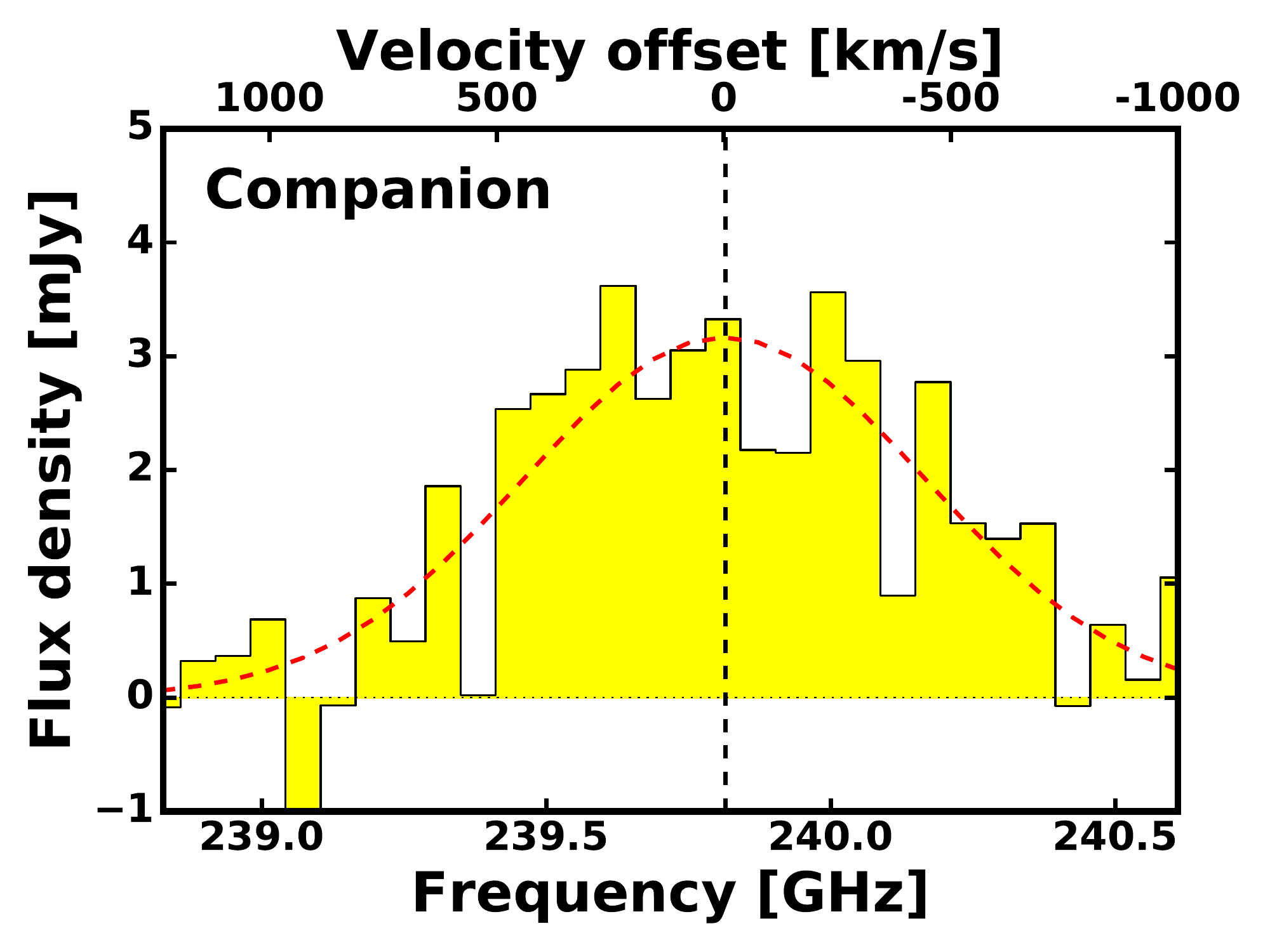}
	\caption{CO(8-7) emission of the system of SMM J04135+10277. \textit{Top:} Integrated intesity contour maps of the quasar and its companion. The contour levels are at $[5,7,10]\times\sigma$, where $\sigma=0.067\rm\,Jy\,beam^{-1}\,km\,s^{-1}$ for the QSO and $\sigma=0.191\rm\,Jy\,beam^{-1}\,km\,s^{-1}$ for the companion. ALMA beams are shown as grey ellipses at the bottom left corner.
\textit{Bottom:} The CO(8-7) spectra of SMM J04135+10277 extracted at the position of the quasar and the companion galaxy, respectively. The spectra are binned to $77\,\rm km\,s^{-1}$ per channel and the continuum is subtracted. The red curves show the single component Gaussian fits to the line profiles. The vertical dashed lines indicate the redshifts of the quasar and the companion obtained from the line fitting of the CO(8-7) emission lines.  The vertical dotted line indicate the optical redshifts of the quasar ($z=2.837$; \citealt{2003A&A...411..343K}). The top-axis shows the relative velocity offset with respect to the fitted redshifts.}
	\label{fig:band6_line}
\end{figure*}

We detect CO(5-4) and CO(8-7) line emission from both the quasar and the companion galaxy (Fig. \ref{fig:band4_line}, \ref{fig:band6_line}). As it was expected from earlier CO(1-0) and CO(3-2) observations, both the CO(5-4) and CO(8-7) emission is dominated by the companion galaxy. 
However, while the quasar remained undetected at low-$J$ CO transitions, it appears bright in our high-$J$ transition observations. This indicates that the host galaxy of the quasar has a significant amount of warm, highly-excited molecular gas.

The line emitting regions of both sources are compact and in the case of the companion, we do not see the elongated shape previously found by \citet{2013ApJ...765L..31R}. 
For further inspection, the spectra of the quasar and companion were extracted from the line data cube. 
We fitted the spectra using Gaussian line profiles in order to determine the line widths and velocity integrated fluxes of the sources.

The CO line profiles of the quasar are narrow and shifted by $\sim700\,\rm km\,s^{-1}$ with respect to the redshift of the quasar derived from rest-frame ultraviolet emission lines  ($z_{\rm UV}=2.837\pm0.003$; \citealt{2003A&A...411..343K}). This is in agreement with previous studies of high-redshift quasars, which found that the broad UV emission lines of quasars tend to be blueshifted by several hundreds of $\rm km\,s^{-1}$ compared to emission lines probing the interstellar medium of the host galaxies, such as the \ion{[C}{ii]} and CO lines \citep[e.g.][]{2011ApJ...730..108R, 2015ApJ...801..123W, 2016ApJ...816...37V, 2017ApJ...836....8T, 2018ApJ...854...97D}. 

In contrast, the CO(5-4) and CO(8-7) line profiles of the companion are very broad, $1127\pm49\,\rm km\,s^{-1}$ and $1038\pm110\,\rm km\,s^{-1}$ respectively, when fitting a single Gaussian profile. There is a descrepancy between the line widths of the different CO transitions: the CO(1-0) and CO(3-2) lines are narrower ($679\pm120\,\rm km\,s^{-1}$ and $765\pm222\,\rm km\,s^{-1}$; \citealt{2013ApJ...765L..31R, 2016ApJ...827...18S}) compared to the higher-$J$ transition lines reported here.
The fitted CO redshifts of the companion shown in Table \ref{tab:table_line} are consistent within the uncertainties of the CO(1-0) redshift reported in \citet{2016ApJ...827...18S} and slightly lower compared to the CO(3-2) redshift presented in \citet{2013ApJ...765L..31R}, which we attribute to the much higher signal-to-noise ratio of our measurements.

\subsection{CO kinematic properties of the companion}
\begin{figure*}
	\centering
	\includegraphics[width=0.98\textwidth]{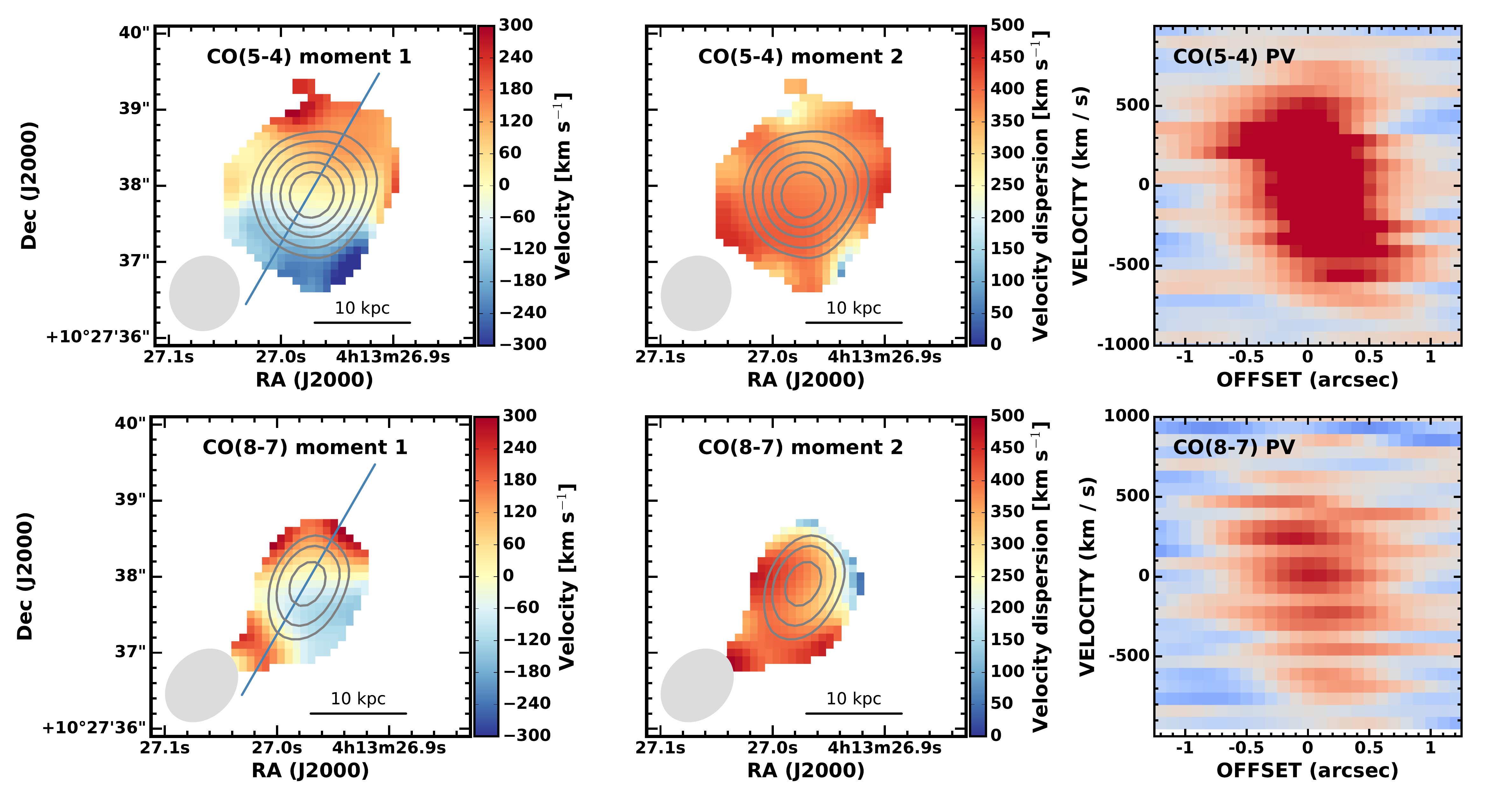}
	\caption{Kinematics of the companion galaxy. \textit{Left:} The velocity map of the companion galaxy. The blue line indicates the position and angle from which the PV diagram was derived. \textit{Middle:} The velocity dispersion map of the companion galaxy. \textit{Right:} The PV-diagram of the companion galaxy. Top and bottom rows show the results for the CO(5-4) and the CO(8-7) line, respectively. Contours trace the integrated line intensity with contour levels same as in Fig. \ref{fig:band4_line} for the CO(5-4) emission and Fig. \ref{fig:band6_line}. for the CO(8-7) emission.}
	\label{fig:velocity}
\end{figure*}

For the companion galaxy, the ALMA observations reveal that the emission of the CO(5-4) and CO(8-7) line is spatially extended over a few beams, though less than suggested in \citet{2013ApJ...765L..31R}. We analysed the intensity weighted velocity field map (moment-1 map), the velocity dispersion map (moment-2 ) and the position-velocity (PV) diagram of the source to characterise the kinematic properties and understand the discrepancy between the linewidth of the low-$J$ and high-$J$ data (Fig. \ref{fig:velocity}). For the PV diagram we chose the angle towards the highest velocity gradient ($-30\degr$) and used the same angle for both transitions. The position and angle from which the PV diagram was derived is indicated as a blue line on Fig. \ref{fig:velocity}. The velocity map and the PV diagram of the CO(8-7) emission line are less well-defined compared to that of the CO(5-4) line, which might be the result of the lower signal-to-noise ratio achieved in band 6.
Both the velocity field maps and the PV diagrams of the high-$J$ transition lines show that the emission is consistent with rotation of a massive and compact system with maximal velocity values of $|V_{\rm max}\,\sin{i}|\gtrsim300\,\rm km\,s^{-1}$ and show no clear signs of outflows, despite the very broad linewidth. 
The velocity dispersion maps of the companion show very high dispersion across the source. A common characteristics of rotating systems is an increase of the velocity dispersion towards the centre. The resolution of the natural weighted image is insufficient to show this, and beam smearing could cancel out this feature. Thus a proper analysis of the kinematics is not possible given that several resolution elements would be needed on either side of a potential rotation curve \citep[e.g.][]{2010MNRAS.401.2113E, 2015MNRAS.451.3021D}. However, in Appendix \ref{sec:galpak} we present a simple kinematic model with rotation using  GalPak$^{\rm3D}$ \citep{2015AJ....150...92B}. 

In contrast to the companion galaxy, the quasar host galaxy does not display a complex line profile and signs of a velocity gradient, hence we do not carry out a similar analysis for this source. In Appendix \ref{sec:qso} we present the moment-1 and moment-2 maps of the quasar.
%
%
%
%
%
\subsection{CO line luminosities and gas masses}
The CO(5-4) and CO(8-7) line luminosities are derived using the following equation from \citet{1997ApJ...478..144S}: $L_{\rm CO}^{'}=3.25\times10^{7}\times S_{\rm CO}dV\times\nu_{\rm obs}^{-2}\times D_{L}^{2}\times(1+z)^{-3}$, where $S_{\rm CO}dV$ is the velocity integrated flux, $\nu_{\rm obs}$ is the observed frequency of the line in GHz and $D_{L}$ is the luminosity distance in Mpc. The derived luminosities of the sources are corrected for the gravitational lensing and are summarised in Table \ref{tab:table_line}.

To estimate the molecular gas mass of the system, we use the CO(5-4) line observations together with the CO(1-0) line observations of \citet{2016ApJ...827...18S}. We take this approach as the molecular gas mass is best estimated from the ground transition of CO and for the companion interferometric CO(1-0) detection is available, while the quasar is only detected in our higher-$J$ ALMA observations. In the further analysis we adopt a conversion factor of $\alpha_{\rm CO}=0.8\,\rm M_{\sun}\,(K\,km\,s^{-1}\,pc^{2})^{-1}$ \citep{1998ApJ...507..615D}, which is typically used for starburst galaxies.

In the case of the companion \citet{2016ApJ...827...18S} reported a line luminosity of $L_{\rm CO(1-0)}^{'}=(8.6\pm1.7)\times10^{10}\, \rm K\,km\,s^{-1}\,pc^{2}$, which converts to a molecular gas mass of $(6.8\pm1.4)\times10^{10}\,\rm M_{\sun}$. This is about a factor of two lower than the molecular gas mass determined by \citet{2013ApJ...765L..31R}, and can be a result of the different resolution of the observations. Regardless of this difference, it is clear that the companion has a massive gas reservoir.

\begin{table}
	\centering
	\caption{Velocity integrated fluxes, line widths and line luminosities of the system of SMM J04135+10277.}
	\label{tab:table_line}
	\begin{tabular}{lcc}
		\hline
		\hline
		&QSO & Companion\\
		\hline
		$SdV_{\rm CO(5-4)}$ ($\rm Jy\,km\,s^{-1}$) &$2.50\pm0.22$ &$8.77\pm0.50$\\
		$SdV_{\rm CO(8-7)}$ ($\rm Jy\,km\,s^{-1}$) & $1.04\pm0.18$ &$3.50\pm0.49$\\
		$\rm FWHM_{\rm CO(5-4)}$ ($\rm km\,s^{-1}$) &$340\pm23$ &$1127\pm49$ \\
		$\rm FWHM_{\rm CO(8-7)}$ ($\rm km\,s^{-1}$) &$278\pm37$ &$1038\pm110$\\
		$z_{\rm CO(5-4)}$ &$2.8460\pm0.0001$ &$2.8438\pm0.0003$\\
		$z_{\rm CO(8-7)}$ & $2.8460\pm0.0002$ &$2.8438\pm0.0006$\\
		$L_{\rm CO(5-4)}^{'}$\textsuperscript{a} ($10^{10}\,\rm K\,km\,s^{-1}\,pc^{2}$ ) &$2.9\pm0.3$ &$8.3\pm0.5$\\
		$L_{\rm CO(8-7)}^{'}$\textsuperscript{a} ($10^{10}\,\rm K\,km\,s^{-1}\,pc^{2}$ ) &$0.5\pm0.8$ &$1.3\pm0.2$\\
		\hline
\end{tabular}
\flushleft
\footnotesize{$^a$ The luminosities are corrected for gravitational lensing magnification: $\mu_{\rm QSO}=1.3$ \citep{2003A&A...411..343K}; $\mu_{\rm Comp.}=1.6$ \citep{2013ApJ...765L..31R}}.\\
\end{table} 

The quasar was not detected by \citet{2016ApJ...827...18S} but a $3\sigma$ upper limit is obtained, yielding a CO(1-0) line luminosity upper limit of $1.9\times10^{10}\, \rm K\,km\,s^{-1}\,pc^{2}$, which converts to a molecular gas mass of $7.4\times10^{9}\,\rm M_{\sun}$.
We also calculated the molecular gas mass from the CO(5-4) emission of the quasar, as this is the lowest-$J$ detection available for this source. We assume that the CO(5-4) transition is thermally excited, based on CO(4-3) observations of high-$z$ type I QSOs, which found a unity $L_{\rm CO(4-3)}^{'}/L_{\rm CO(1-0)}^{'}$ ratio \citep[e.g.][]{2006ApJ...650..604R, 2007A&A...467..955W}.
With this assumption the molecular gas mass of the quasar is $(2.3\pm0.2)\times10^{10}\,\rm M_{\sun}$. These results suggest that there is a significant amount of warm gas in the quasar and it cannot be considered gas-poor.
\subsection{\texttt{RADEX} modelling of the CO spectra}
To put constraints on the excitation properties of SMM J04135+10277, we modelled the SLED of both sources using our ALMA detections combined with CO(1-0) and CO(3-2) line observations from the literature \citep{2013ApJ...765L..31R, 2016ApJ...827...18S}. We used the non-LTE molecular radiative transfer code \texttt{RADEX} \citep{2007A&A...468..627V} to calculate the line intensities of CO from transitions $J_{\rm up}=1$ to $J_{\rm up}=8$, within a wide range of physical parameters: kinetic temperature ($T_{\rm kin}$), number density of molecular hydrogen ($n_{\rm H_{2}}$), and column density of CO ($N_{\rm CO}$). During each run we set the background radiation temperature to the value of the cosmic microwave background at $z=2.84$, adopted spherical geometry and assumed that the only collisional partner of CO is molecular hydrogen. As the linewidths of the low- and high-$J$ CO lines are different, we use an average value during the \texttt{RADEX} modelling. The \texttt{RADEX} code only considers photodissociation regions and does not include mechanical and X-ray heating. We compared our observations with the models using $\chi^{2}$-analysis.
In the first run we varied all three physical parameters in case of each source but as most of the models converged to the same CO column density, we fixed $N_{\rm CO}$ and continued with fitting only $T_{\rm kin}$ and $n_{\rm H_{2}}$. 

During the \texttt{RADEX} modelling of the quasar we set the line width of $300\,\rm km\,s^{-1}$, fixed $N_{\rm CO}$ to $10^{16}\,\rm cm^{-2}$ and run \texttt{RADEX} on a grid of $T_{\rm kin}=[10-400]\,\rm\,K$ and $\rm log(\it n_{\rm H_{2}})=[\rm 2-7]\,\rm\,cm^{-3}$. It is important to note, that as we only have two detections for the quasar and two upper limits, we do not expect to arrive to a straightforward conclusion about the excitation properties of this source. However, by using the available data we can narrow down the possible temperature and density ranges. 

\begin{figure*}
	\centering
	\includegraphics[width=1.\columnwidth]{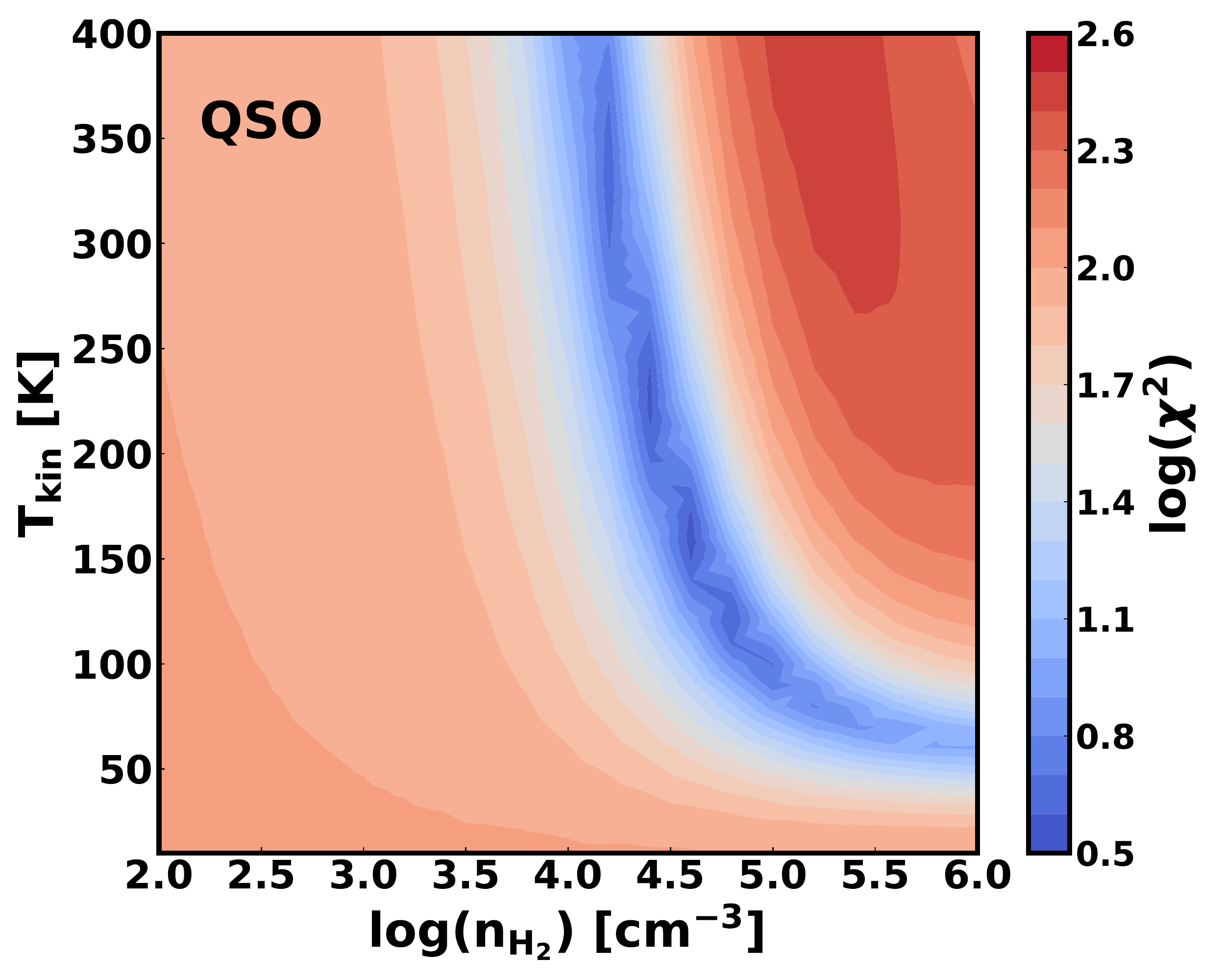}
	\includegraphics[width=1.\columnwidth]{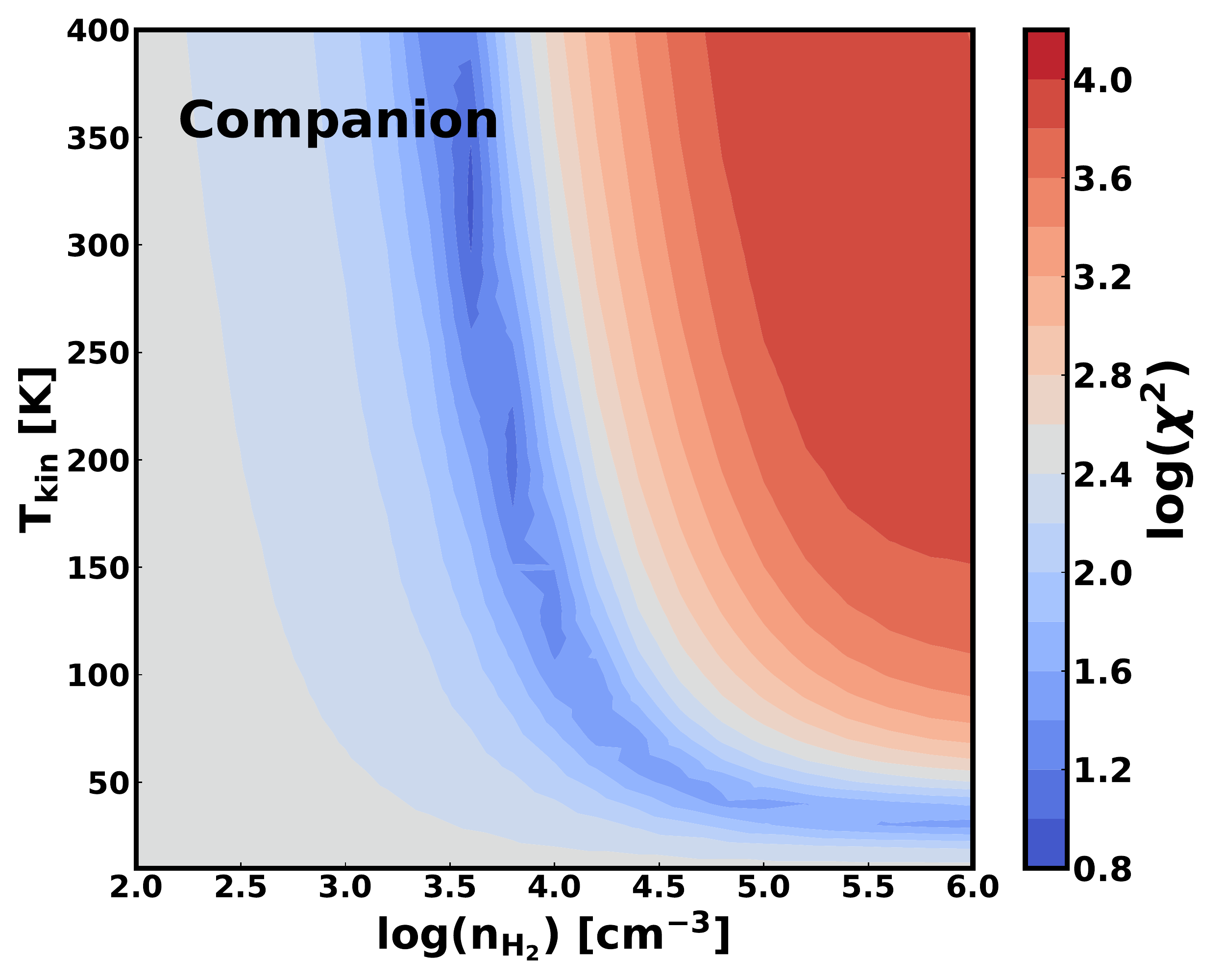}
	\caption{$\chi^{2}$ contour plots of SMM J04135+10277. The contour plots trace the $\chi^{2}$ value of the \texttt{RADEX} CO models as a function of molecular hydrogen density and temperature. The left panel shows the fitting results of the quasar, the right panels shows the same for the companion galaxy.}
	\label{fig:chi2}
\end{figure*}

\begin{figure*}
	\centering
	\includegraphics[width=1.\columnwidth]{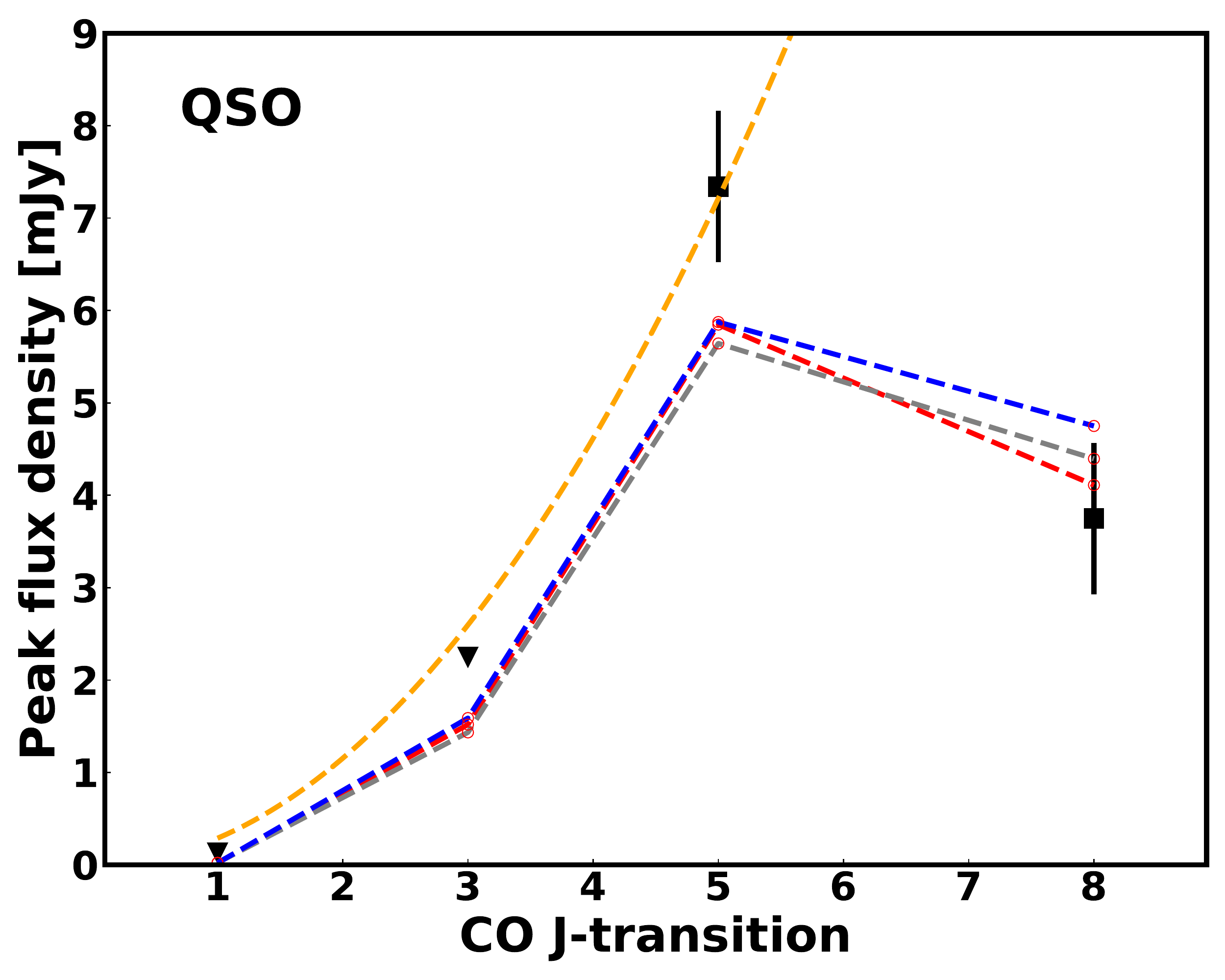}
	\includegraphics[width=1.\columnwidth]{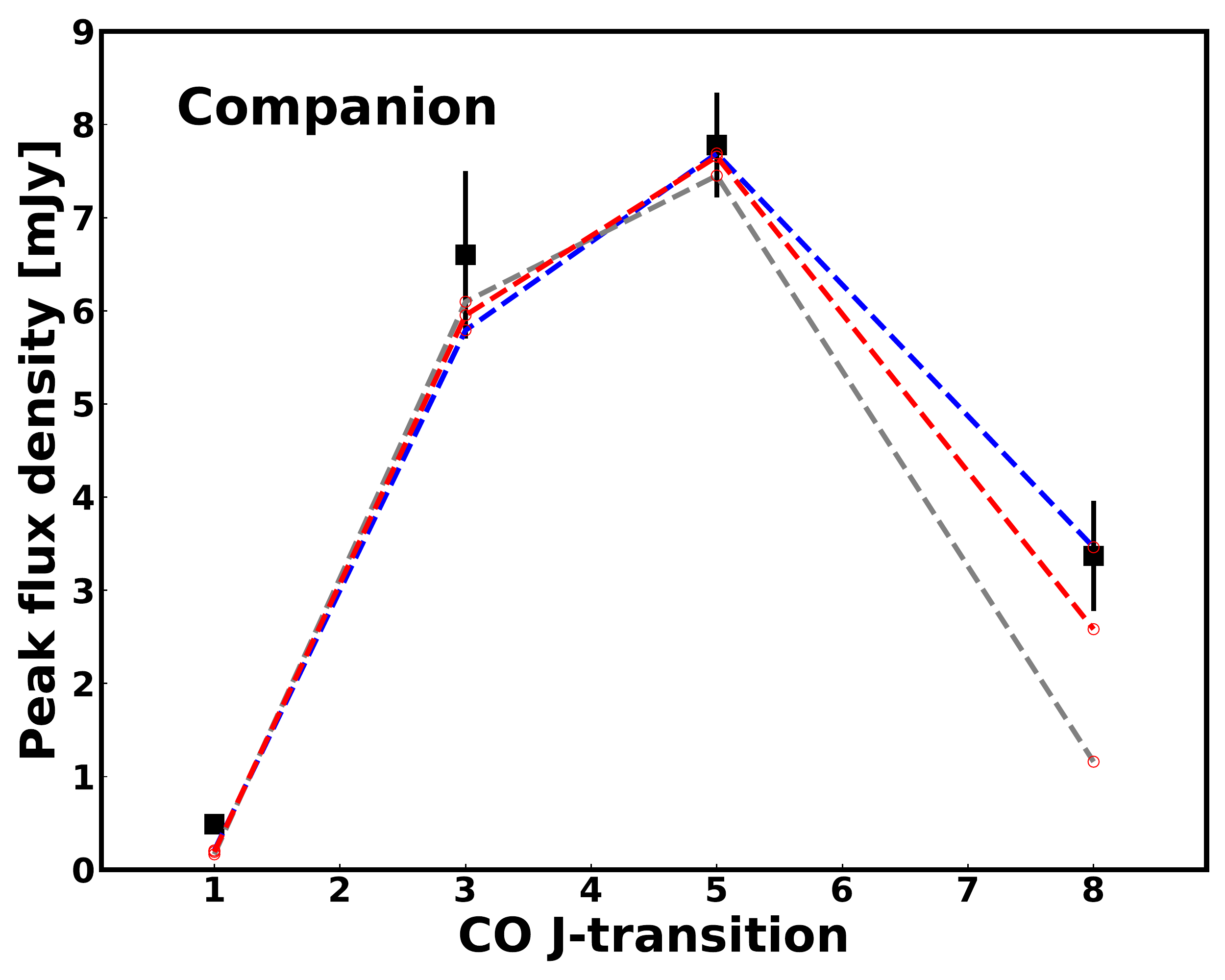}
	\caption{Spectral line energy distribution of SMM J04135+10277. The left panel shows the \texttt{RADEX} SLED fitting of the quasar. We selected three different temperature models with low $\chi^{2}$ value. The blue, red and grey curves correspond to models with a $T_{\rm kin}=[360, 160, 100]\,\rm K$ and $\rm log(\it n)=[\rm 4.2, 4.6, 5.0]\,\rm cm^{-3}$, respectively. The yellow dashed line indicates the $J^2$ scaling, the expected scaling of the fluxes in the Rayleigh-Jeans limit and in case of thermal excitation. The right panel shows the same for the companion galaxy. The blue, red and grey curves correspond to models with a $T_{\rm kin}=[530, 300, 110]\,\rm K$ and $\rm log(\it n)=[\rm 3.4, 3.6, 4.0]\,\rm cm^{-3}$, respectively.}
	\label{fig:sled}
\end{figure*}

Figure \ref{fig:chi2} shows how the $\chi^{2}$ value of the models vary as a function of temperature and density. 
The upper-$J$ transitions of the quasar can be fitted by models with a continous range of kinetic temperatures of $60-400\,\rm K$ and molecular hydrogen density between $10^{4}-10^{6.2}\,\rm cm^{-3}$.
Models with the lowest $\chi^{2}$ values are outside of the typical temperature range for quasars ($T_{\rm kin}=40-60\,\rm K$; \citealt{2006ApJ...650..604R,2009ApJ...703.1338R,2013ARA&A..51..105C}) and the corresponding densities are much higher compared to the typical quasar value ($10^{3.6}-10^{4.3}\,\rm cm^{-3}$). This might be due to the lack of data points to produce a reasonable fit but it could also suggest the presence of an extremly compact region of highly excited gas close to the central AGN. 
In contrast, there are models with low $\chi^{2}$ value and densities within the typical range for quasars but their temperature is not well-constrained.
In order to illustrate how the \texttt{RADEX} models fit the observations, on Figure \ref{fig:sled} we plot three model SLEDs with low $\chi^{2}$ value but with different temperatures and densities. In addition, we calculated the molecular gas mass of the quasar using the CO(1-0) line intensity of the the best-fit models, which gives a mass range of $(1.2-1.8)\times 10^{9}\,\rm M_{\sun}$. The \texttt{RADEX} estimated gas mass is lower compared to the gas mass derived from the CO(5-4) transition and the CO(1-0) upper limit.

For the modelling of the companion galaxy we used a line width of $850\,\rm km\,s^{-1}$, fixed $N_{\rm CO}$ to $10^{17}\,\rm cm^{-2}$ and run \texttt{RADEX} on a grid of $T_{\rm kin}=[10-400]\,\rm\,K$ and $\rm log(\it n_{\rm H_{2}})=[\rm 2-7]\,\rm\,cm^{-3}$. 
From the $\chi^{2}$-analysis we found that models with low $\chi^{2}$ value give a consistent estimate for the molecular hydrogen density, between $10^{3}-10^{4}\,\rm cm^{-3}$, but can have temperatures ranging from $110\,\rm K$ to $400\,\rm K$ (Fig.\ref{fig:chi2}). 
While the temperatures of the best-fit models are not well constrained, they could narrow down the possible molecular hydrogen density range of the companion, and the yielded range is within the typical values found for SMGs ($10^{2.7}-10^{3.5}\,\rm cm^{-3}$; \citealt{2013ARA&A..51..105C}). Using the CO(1-0) line intensity of the best-fit models, the molecular gas mass of the companion is $(2.9-4.5)\times 10^{10}\,\rm M_{\sun}$, which is fairly close to the gas mass derived from the CO(1-0) transition.
On Figure \ref{fig:sled} we show three models SLEDs which have low $\chi^{2}$ value, but have very different kinetic temperatures. 
Based on Figure \ref{fig:sled} it is clear that even the best-fit \texttt{RADEX} models cannot fit well the data and the temperature cannot be simply constrained.

\section{Discussion}
\label{sec:discussion}
\subsection{FIR emission of SMM J04135+10277}
Thanks to the high-resolution of ALMA we were able the resolve the system of SMM J04135 and detect FIR emission from both the quasar and its companion. However, fitting the FIR SED of the sources is still challenging, as the ALMA observations only trace emission at the Rayleigh-Jeans side of the FIR SED and adequate observations around the peak of the FIR SED are lacking. It has been shown that the wavelength of the rest-frame SED peak can be used as a proxy for dust temperature and it decreases with increasing total infrared luminosity but it is dependent on the emissivity and opacity of the used model \citep{2013ApJ...778..131L}. Furthermore, the dust temperature increases with infrared luminosity \citep{2012ApJ...761..140C}. Thus, the infrared luminosities and SFRs of the quasar and the companion estimated from the \textsc{MR-MOOSE} models can be considered as lower limits, and a better sampling around the peak of the SEDs is required to get the whole picture.

Another important note about the FIR SED of the quasar is the possible contribution of the AGN to the total FIR emission. Several studies addressed this question with contradicting results and this topic is still highly debated\citep[e.g.][]{2006ApJ...649...79S, 2014ApJ...785..154L, 2015A&A...579A..60S, 2016A&A...591A.136L, 2017MNRAS.465.1401S}. What can be assumed based on these studies is that the most luminous AGNs can have a significant contribution to IR emission, mostly at near- and mid-IR wavelength originating from the dusty torus component of AGNs. The main obstacle, which needs to be eliminated is the resolution factor, as many studies use large-beam \textit{Herschel} observations, thus potentially detect dust emission from blended sources. Since in our case the dust emission of SMM J04135 is only resolved in the ALMA bands, it is very difficult to estimate the AGN contribution at lower wavelength, in the SCUBA bands in particular.

\subsection{Rotation of a massive system}
The observed CO(5-4) and CO(8-7) lines of the companion are very broad, with about a $1000\,\rm km\,s^{-1}$ FWHM in each band. These values are much higher compared to lower-$J$ transitions reported in the literature \citep{2013ApJ...765L..31R, 2016ApJ...827...18S}. Looking at other high-$z$ sources, only a few submm galaxies and quasars have been reported to have such broad CO lines \citep{2003ApJ...584..633G, 2003ApJ...597L.113N, 2005MNRAS.359.1165G, 2008MNRAS.389...45C, 2011A&A...533A..20P,2018ApJ...860...87F}, making the companion galaxy special.

In order to find the reason of observing such a broad line, we obtained the kinematics of the companion galaxy through velocity maps, velocity dispersion maps and PV diagrams. Based on our analysis we found no features characteristic to molecular outflows and the velocity map of the companion suggests rotation of a massive and compact galaxy with maximal velocity values of $|V_{\rm max}\,\sin i|\gtrsim300\,\rm km\,s^{-1}$. The velocity dispersion map shows an almost uniform dispersion all over the companion galaxy, which could be an indication of spatially unresolved rotation \citep{2011A&A...528A..88G}.

As we did not find signs of outflows, we consider two possible explanations for the origin of the broad CO lines, which can happen separately or at the same time. The broad line could be the result of not having sufficient resolution to map the companion galaxy and observe the velocity gradients both at large and small scales. Beam smearing has been known as one of the main effects causing broad lines and increasing the observed dispersion of galaxies \citep[e.g.][]{2002ASPC..275..217T, 2013ApJ...767..104N, 2014A&A...565A..59D, 2019MNRAS.483..249H}. At high-$z$ this effect can be even more prominent. As the ALMA beam sizes of our observations are comparable of the CO size of the companion galaxy, this effect is a plausible explanation for observing such broad lines.

Even if the broad line is consistent with rotation, another reason why the observed CO lines are broad could be turbulence, as it has been found in case of recent studies of high-$z$ galaxies \citep[e.g.][]{2011A&A...528A..88G, 2014MNRAS.443.3780W}. Turbulence could arise from several components, such as having a hidden AGN or from a late stage merger event in the companion galaxy, where the merging galaxies are in coalescence. Given the high SFR of the companion galaxy, the supernova rate might be also high, thus the kinematic energy released by supernovae could also account for  part of the turbulence.
To investigate these scenarios, higher resolution observation are required, preferably tracing even higher-$J$ transitions as well.

\subsection{CO excitation properties of SMM J04135+10277}
The detection of CO(5-4) and CO(8-7) line emission in both sources complemented with low-$J$ CO observations provides the opportunity to infer the excitation properties of this system. However, there are still some limitations to get the whole picture about the excitation conditions, such as having too few data points to fit the SLED.
In case of the quasar we only have detections in the ALMA bands and upper limits for the CO(1-0) and CO(3-2) emission. This of course affects the $\chi^{2}$-analysis and thus the \texttt{RADEX} models with low $\chi^{2}$ value yield very different solutions.

In addition to lacking enough data points, another reason we cannot fit well the observations might be the model itself. By using a model with a more complex radiation field and geometry other than spherical could change the outcome and better fit the data. It has been shown that high-$J$ transitions starting from $J=7$ are better fitted with models including  X-ray dominated regions (XDR; \citealt{2007A&A...461..793M}). Such models could be especially relevant for the quasar, where the central compact region maybe better modelled by an XDR, while the more diffuse gas associated to star formation is better modelled by a PDR \citep{2005A&A...436..397M}.
However, models including XDRs have more fitted parameters than the \texttt{RADEX} model used in this paper, and thus could be done provided the SLED was better sampled including more data for $J>8$ transitions.

The detection of highly excited molecular gas in the host galaxy of the quasar shows that one has to be careful when interpreting non-detections of low-$J$ CO transitions in such sources.
For example, \citet{2017MNRAS.468.4205K} presented CO(2-1) observations of $z\sim1.5$ quasars, with a detection rate of only 30\%.  Based on our results we note that the low detection rate could be related to the compactness of the CO emitting regions and also because the excitation conditions favour the excitation of high-$J$ transitions, rather than the absence of a massive molecular gas reservoir.

The situation is not less complicated when we look at the SLED of the companion galaxy. While we have four CO detections, it is clear that the companion cannot be fitted with a single phase component. In case of SMGs, it has been already demonstrated that their SLED is often best modelled with a combination of an extended, diffuse component and a compact component with higher excitation \citep{2011ApJ...739L..31R, 2010ApJ...723.1139H, 2013ApJ...768...91H}. However, by fitting two separate components to the low- and high-$J$ CO transitions we encounter the same problem, such as having two observations and two fitted parameters. Therefore, we cannot infer both the temperature and density of the sources.

\subsection{Quasar--star-forming companion systems}
In recent years, many high-redshift quasars have been observed with ALMA taking advantage of its sensitivity and high-resolution.While some of these studies, tracing dust and molecular gas emission, found vigorous star formation and huge gas reservoirs in the host galaxies of the quasars \citep[e.g.][]{2013ApJ...773...44W, 2017MNRAS.465.4390B}, the number of detected quasar--star-forming companion galaxy systems has also been growing \citep[e.g.][]{2017MNRAS.465.4390B, 2017A&A...605A.105C, 2017Natur.545..457D, 2017ApJ...836....8T}.
This is in agreement with our simulation results reported in \citet{2017A&A...597A.123F}, where we used \textsc{GALFORM} \citep{2014MNRAS.439..264G, 2014MNRAS.440..920L}, a galaxy formation and evolution model to investigate the expected frequency of finding systems similar to SMM J04135+10277. According to the simulations, at a distance of $<350\,\rm kpc$, 33\% of the simulated quasar sample have a companion galaxy ($M_{\star}>10^{8}\,\rm M_{\sun}$) and 2.4\% have bright companions with a SFR $>100\,\rm M_{\sun}\,yr^{-1}$. 

However, the system of SMM J04135+10277 seems to differ from the other detected quasar--companion pairs, as the dust emission and the SFR is dominated by the companion\footnote{We note that while SFRs of the sources are comparable based on the \textsc{MR-MOOSE} analysis, which makes SMM J04135+10277 more similar to other AGN--companion systems, given the few detections a complete SED fitting is challenging. Thus we treat the \textsc{MR-MOOSE} values as lower limits to the SFR.}. This could be due to the different redshifts of the studied sources, but it might indicate that the companion galaxy of SMM J04135+10277 harbours a hidden AGN, which also has a contribution to the FIR emission. SMGs harbouring hidden AGNs is not a new concept, and in some sources, X-ray observations have already revealed obscured AGNs \citep[e.g.][]{2003AJ....125..383A, 2005ApJ...635..853B}. Although both \textit{Chandra} and  \textit{XMM-Newton} have observed the field, the observations targeted the foreground X-ray bright galaxy cluster making it difficult to resolve and disentangle emission from the QSO and companion.

Another possible explanation why we do not see many more systems reminiscent of SMM J04135+10277 is the time scale issue: it could be that we captured the system of SMM J04135+10277 at a special time of its evolution, with the companion galaxy having an intense star-forming phase and the quasar being active. Given the short duration of the starburst phase ($<10^{8}\,\rm yr$) and quasar lifetime ($<10^{8.5}\,\rm yr$, \citealt{2008A&A...492...31D, 2008ApJS..175..356H}), it might be that we simply missed this window of time in case of other known systems and observed them in a later stage of their evolution. 
\section{Conclusions}
\label{sec:conclusion}
We have presented ALMA observations of the dust continuum, CO(5-4) and CO(8-7) line emission of the quasar--companion galaxy system SMM J04135+10277 ($z=2.84$). Compared to previous studies of the system using large beam observations, we resolve the continuum emission and detect dust emission associated with both sources. Based on the ALMA continuum data, the dust emission is dominated by the star-forming companion galaxy but the quasar has a non-negligible contribution of 25\% to the total emission. We fitted the SED of the sources using the SED fitting code \textsc{MR-MOOSE}, which is designed to treat upper limits and blended sources using a Bayesian approach. Based on the SED fitting the dust temperature of the quasar is higher compared to that of the companion, resulting in similar FIR luminosities for each source.

As the companion galaxy shows a very broad line profile in both CO transitions ($\sim 1000\,\rm km\,s^{-1}$), we studied the kinematics of the galaxy. The ALMA results show signs of rotation, however, in the absence of high resolution observations a proper analysis of the kinematics is not possible and the observed line width of the companion is possibly affected by beam smearing.

While previous observations of the CO(1-0) and CO(3-2) transitions only detected molecular line emission associated with the companion galaxy, we detect a significant molecular gas reservoir in each source. The estimated  molecular gas mass of the companion galaxy and the quasar host is $\sim7\times10^{10}\,\rm M_{\sun}$ and $\sim0.7-2.3\times10^{10}\,\rm M_{\sun}$, assuming a conversion factor of $\alpha_{\rm CO}=0.8\,\rm M_{\sun}\,(K\,km\,s^{-1}\,pc^{2})^{-1}$. In the light of our observations, it is clear that the quasar is not gas-poor as it was suggested by previous studies and has a significant molecular gas mass, only visible at higher frequencies due to excitation.

Using the results of low-$J$ CO transitions observations found in the literature and our ALMA CO(5-4) and CO(8-7) detections, we model the SLED of each source using the radiative transfer code \texttt{RADEX}. In the case of the quasar we only have limited amount of detections, thus the model cannot constrain well the excitation properties of the source. However, models with low $\chi^{2}$ value could narrow down the possible temperature and density range.
For the companion the \texttt{RADEX} models cannot put constraints on the temperature but give a consistent estimate for the molecular hydrogen density.
The reason of this might be the limited number of observations and the simplicity of the model.

Finally, we compared the case of SMM J04135+10277 to other quasar--companion systems observed by ALMA. In comparison, SMM J04135+10277 stands out from similar systems as the quasar does not dominate the dust emission, while still having a significant amount of molecular gas.\\

\section*{Acknowledgements}
We thank the anonymous reviewer for the constructive report that helped us to improve the manuscript. J.F. would like to thank Sabine K\"{o}nig for her help and valuable expertise of ALMA data reduction. J.F. would like to thank Alessandro Romeo, John H. Black and Susanne Aalto  for the useful discussions. J.F. acknowledges support from the Nordic ALMA Regional Centre (ARC) node based at Onsala Space Observatory. The Nordic ARC node is funded through Swedish Research Council grant No 2017-00648.
J.F. and K.K. acknowledge support from the Knut and Alice Wallenberg Foundation. K.K. acknowledges support from the Swedish Research Council. L.F. acknowledges the support from the National Natural Science Foundation of China (NSFC, Grant Nos. 11822303 and 11773020).
This paper makes use of the following ALMA data: ADS/JAO.ALMA\#2015.1.00661.S. ALMA is a partnership of ESO (representing its member states), NSF (USA) and NINS (Japan), together with NRC (Canada) and NSC and ASIAA (Taiwan) and KASI (Republic of Korea), in cooperation with the Republic of Chile. The Joint ALMA Observatory is operated by ESO, AUI/NRAO and NAOJ.




\bibliographystyle{mnras}
\bibliography{fogasy_alma_qso_ref1}

\appendix

\section{GalPak$^{\rm3D}$ modelling of the companion galaxy}
\label{sec:galpak}
As described in the main text, the angular resolution of the ALMA data does not allow us for a detailed analysis of the kinematics of the companion galaxy. However, given the velocity gradient seen in both the CO(5-4) and CO(8-7) emisson, it is possible that at least a fraction of the data could be represented by a rotating disk. Thus we carry out a simple analysis of the kinematics, assuming a rotating system, using GalPak$^{\rm3D}$ \citep{2015AJ....150...92B}.
%
%
GalPak$^{\rm3D}$ is a Bayesian parametric tool to constrain galaxy parameters directly from three-dimensional data cubes \citep{2015AJ....150...92B}. 
As an input to GalPak$^{\rm3D}$, we assume a rotating disk with an exponential brightness distribution and with rotational velocity profile described with an arctan profile $v(r) \propto V_{\rm max}\,{\rm arctan}(r/r_{\rm t})$, where $V_{\rm max}$ is a maximum circular velocity and $r_t$ is the turnover radius \citep{2015AJ....150...92B}. Using the natural weighted cube, we ran the fitting several times with improved starting values for the Bayesian fitting.
We note that we use the default settings as the resolution of the data is insufficient to distinguish between different models and the main goal of this simple analysis is to illustrate the possibility of the companion being a rotating system.

GalPak$^{\rm3D}$ models the companion as a compact source with an effective radius of $3.0\pm0.04\,\rm kpc$, velocity dispersion of $285\pm4\,\rm km\,s^{-1}$, maximum velocity of $955\pm5\,\rm km\,s^{-1}$, and inclination of $28.2\degr\pm0.4\degr$, where the uncertainties reflect the $1\sigma$ estimates from the Bayesian fitting itself.  
While we used the natural weighted data, we also repeated the fitting using the data cubes that were imaged using Briggs weighting with a robust parameter of 0.5 and -1.0 (a lower robust parameter means higher angular resolution but worse sensitivity in the data). The results of these runs are consistent with that of the fitting to the natural weighted data, and the effective radius is in the range of $2-3\,\rm kpc$, the velocity dispersion is in the range of $200-300\,\rm km\,s^{-1}$, the maximum velocity is in the range of $900-960\,\rm km\,s^{-1}$, and the inclination in the range of $25\degr-29\degr$ (these ranges are based on the highest and lowest 95\% confidence intervals of the resulting parameters for fitting to all three resolutions). 
Most important is to note that without significantly better angular resolution, it is not possible to properly constrain the inclination and make a proper constraint on the maximum velocity. However, while the maximum velocity of $\sim900\,\rm km\,s^{-1}$ most likely reflects the effect of beam smearing, finding a large velocity dispersion and/or large rotational velocity is not surprising given the compactness and large gas mass of the companion galaxy.

\begin{figure*}
	\centering
	\includegraphics[width=1.\textwidth]{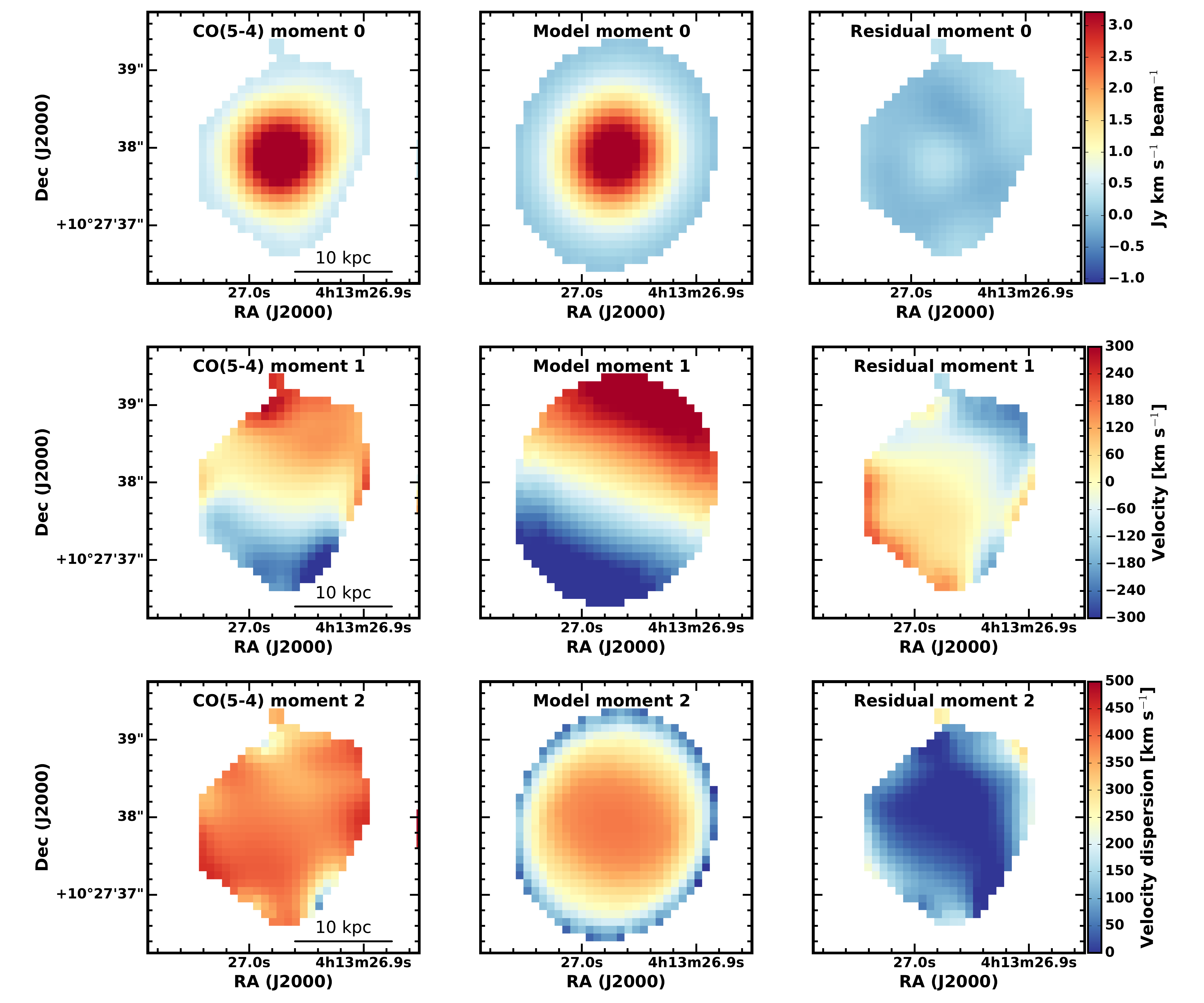}
	\caption{CO(5-4) kinematics of the companion galaxy using the natural weighted data. From left to right the panels show the derived moment maps of the CO(5-4) data, the GalPak$^{\rm3D}$ model and the residual image. The top panels show the integrated intesity maps, the middle panels show the velocity maps, the bottom panels show the velocity dispersion maps.}
	\label{fig:galpak_comp}
\end{figure*}

On  Fig. \ref{fig:galpak_comp}, we show the moment 0, 1,  and 2 maps of the data together with that of the model. In the moment 0 and 2 maps, we see that the model is consistent with the data, but the moment 1 map shows some velocity residuals. While it is possible to make a model description of a rotating system for the companion galaxy, high angular resolution data are necessary in order to fully characterise the kinematics.

\section{Kinematics of the quasar}
\label{sec:qso}
\begin{figure*}
	\centering
	\includegraphics[width=1.\textwidth]{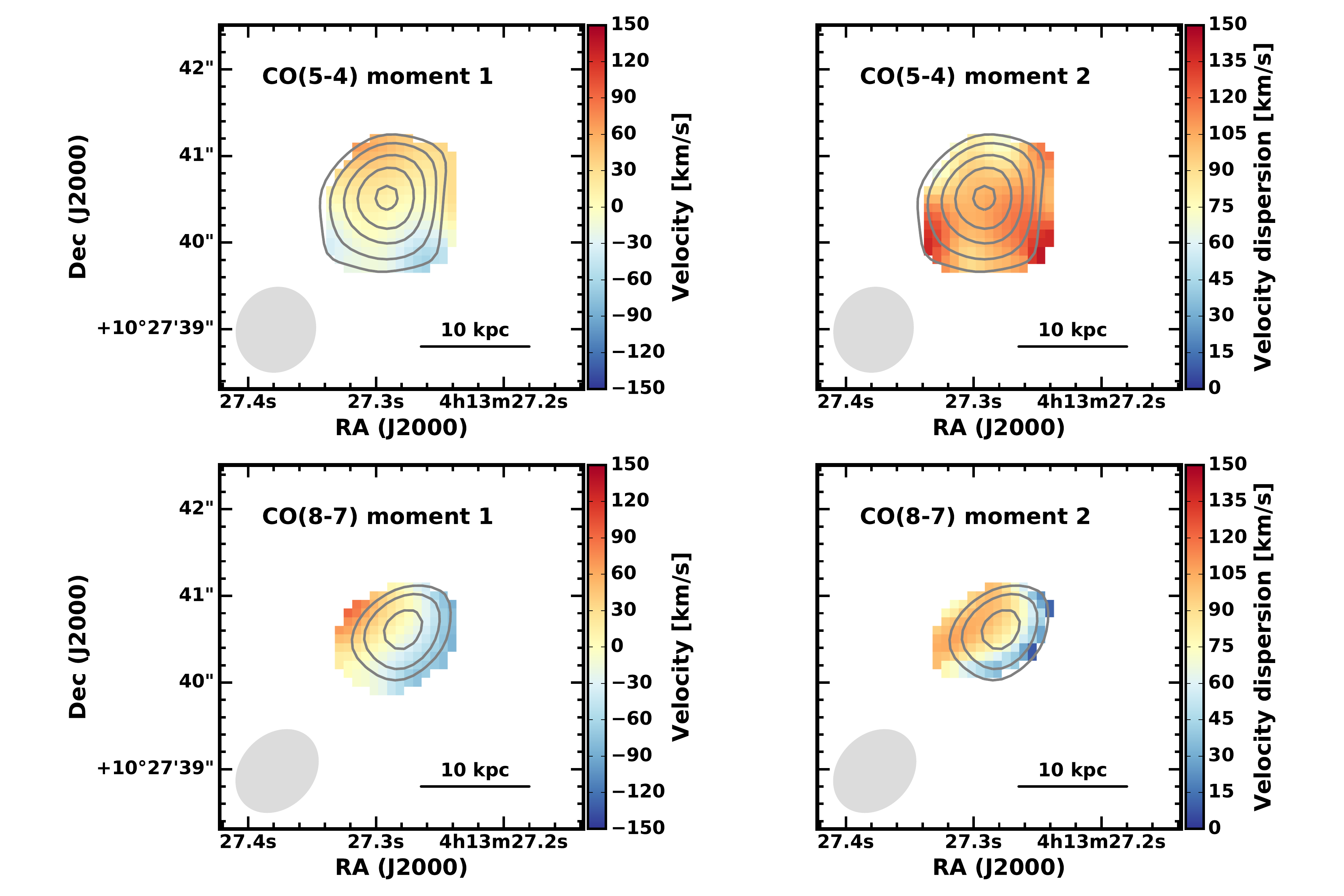}
	\caption{Moment maps of the quasar. \textit{Left:} The velocity map of the quasar. \textit{Right:} The velocity dispersion map of the quasar. Top and bottom rows show the results for the CO(5-4) and the CO(8-7) line, respectively. Contours trace the integrated line intensity with contour levels same as in Fig. \ref{fig:band4_line} for the CO(5-4) emission and Fig. \ref{fig:band6_line}. for the CO(8-7) emission. 
ALMA beams are shown as grey ellipses at the bottom left corner.}
	\label{fig:qso}
\end{figure*}

As the quasar is very compact and unresolved in our observations and there is no signature of a velocity gradient, we do not carry out a GalPak$^{\rm3D}$ analysis, as was done for the companion. On  Fig. \ref{fig:qso} we show the velocity (moment-1) and velocity dispersion (moment-2) maps of the quasar in the case of both transitions. All maps have been cut to the $3\sigma$ level of the moment-0 map.

\bsp	
\label{lastpage}
\end{document}